\documentclass[longauth]{aa}

\usepackage{txfonts}			

\usepackage[utf8]{inputenc}	
\usepackage{amsmath}			
\usepackage{nccmath}			

\usepackage[english]{babel}	
\usepackage{xcolor}			
\usepackage{graphicx}			
\usepackage{booktabs}			
\usepackage{gensymb}			

\usepackage[
colorlinks = true,
anchorcolor = blue,
linkcolor = blue,
urlcolor  = blue,
citecolor = blue,
]{hyperref}

\usepackage[colorinlistoftodos]{todonotes}
\usepackage{textcomp}
\usepackage{orcidlink}
\usepackage{longtable}
\usepackage{array}

\newcolumntype{P}[1]{>{\raggedright\arraybackslash}p{#1}}

\newcommand{\SM}{1995~SM$_\textrm{55}$}

\let\oldsim\sim
\renewcommand{\sim}{{\oldsim}}

\begin{document}
\title{A high geometric albedo and small size of the Haumea cluster member (24835) \SM{} from a stellar occultation and photometric observations} 
\titlerunning{Physical properties of TNO \SM{}}

\author{
J.~L.~Ortiz\inst{1}\orcidlink{0000-0002-8690-2413}
\and
N.~Morales\inst{1}\orcidlink{0000-0003-0419-1599}
\and
B.~Sicardy\inst{2}\orcidlink{0000-0003-1995-0842}
\and
F.~L.~Rommel\inst{3}\orcidlink{0000-0002-6085-3182}
\and
F.~Braga-Ribas\inst{4,5,6}\orcidlink{0000-0003-2311-2438}
\and
Y.~Kilic\inst{1,6}\orcidlink{0000-0001-8641-0796}
\and
E.~Fern\'andez-Valenzuela\inst{3}\orcidlink{0000-0003-2132-7769}
\and
J.~L.~Rizos\inst{1}\orcidlink{0000-0002-9789-1203}
\and
B.~Morgado\inst{7}\orcidlink{0000-0003-0088-1808}
\and
L.~Catani\inst{7}
% ...
}

\authorrunning{Ortiz et al.}

\institute{
Instituto de Astrofísica de Andalucía, IAA-CSIC, Glorieta de la Astronomía s/n, 18008 Granada, Spain \and 
LTE, Observatoire de Paris, Universit'e PSL, Sorbonne Universit'e, Universit'e de Lille, LNE, CNRS 61 Avenue de l'Observatoire, 75014 Paris, France
\and
Florida Space Institute (FSI) - University of Central Florida (UCF), Partnership I, Research Parkway, 32826 Orlando, United States of America
\and
Federal University of Technology -- Paran\'a (PPGFA/UTFPR-Curitiba), Av. Sete de Setembro, 3165, Curitiba -- PR, Brazil
\and
Laboratório Interinstitucional de e-Astronomia - LIneA, Av. Pastor Martin Luther King Jr 126, 20765-000, Rio de Janeiro, RJ, Brazil
\and
LIRA, CNRS UMR8254, Observatoire de Paris, Université PSL, Sorbonne Université, Université Paris Cité, CY Cergy Paris Université, 92190 Meudon, France
\and
Federal University of Rio de Janeiro - Observatory of Valongo, Rio de Janeiro, Brazil
% ...
}

\date{Received dd mmm, 2024; accepted dd mmm, 2024}

\abstract
{Trans-Neptunian objects (TNOs) are thought to be some of the most ancient and primitive bodies in our solar system. Understanding their basic physical properties is crucial to unraveling their origin and the evolution of the outer solar system beyond Neptune. Stellar occultations are a highly effective and sensitive method of studying these distant and faint objects, allowing us to gather essential information about their physical characteristics.  (24835) \SM{} is one of the few members of
the Haumea orbital cluster, and therefore is an especially relevant body to study within the TNO population.}
{ The main objectives of the present work  are to determine the projected size, absolute magnitude, and geometric albedo of \SM{} and to analyze the results compared to Haumea.}
{ We predicted a stellar occultation by this TNO for 25 February 2024, carried out a specific campaign to observe the occultation, and derived the projected  size and shape from the occultation observations using an elliptical fit to the occultation chords. We also analyzed a large set of photometric observations of (24835) \SM{} to obtain the absolute magnitude and the rotational period. Finally, we combined these results to derive the geometric albedo of this TNO.}
{ The occultation was successfully detected from 7 instruments located at five different sites and was negative from 33 other sites. Using an elliptical fit to the occultation chords, we obtained the limb of (24835) \SM{} during the occultation, resulting in an ellipse with semi-axes $(104.3 \pm 0.4) \times (83.5 \pm 0.5)$~km. The area-equivalent diameter for this ellipse is $D_\mathrm{eq,A} = 186.7 \pm 1.8$~km. This is smaller than the upper limit of 250 km from Herschel Observatory thermal data. From our photometric observations, we derived an absolute magnitude $H_V = 4.55 \pm 0.03$, a phase slope parameter of $0.04 \pm 0.02 $ mag/deg and a $V-R = 0.37 \pm 0.05$ value. The rotational variability has a maximum peak-to-valley amplitude $\Delta m = 0.05 $ mag, but we could not derive an unambiguous rotational period. Combining the projected size from the occultation with the absolute photometry, we obtain a geometric albedo in the V band of $p_V = 0.80 \pm 0.04$ for \SM{}. This value is remarkably high for a TNO and somewhat higher than that of Haumea, but consistent with the concept that \SM{} is a member of the orbital cluster of Haumea.}
{}


\maketitle

\section{Introduction}\label{sec:introduction}

Trans-Neptunian objects (TNOs), defined as celestial bodies with a semi-major axis exceeding Neptune's, along with Centaurs (thought to be in a transitional phase between TNOs and Jupiter-family comets \citep[e.g.,][]{Horner2004,Sarid2019}), are considered some of the least altered objects since the solar system's genesis. Their preserved primordial characteristics and materials offer a unique window into the early stages of solar system formation and evolution. Consequently, investigating their physical and dynamical attributes is a potent method to understand these formative periods. Currently, over 5000 TNOs and Centaurs have been identified, though their total population is anticipated to surpass that of asteroids in the main asteroid belt.

Due to their considerable distance from the Sun, TNOs present a challenge for optical observation, as the brightness of a solar system body diminishes with the fourth power of its heliocentric distance. The vast majority of these objects typically exhibit magnitudes fainter than 20 in visible light. Furthermore, with average surface temperatures between approximately 30-40 ~K, their thermal emission peaks in the far-IR, a spectral region significantly attenuated by Earth's atmosphere. Aside from a few exceptionally large TNOs, determining radiometric sizes for this population generally requires observations with ALMA or space telescopes, similar to the approach taken by the ESA Herschel mission's "TNOs are Cool" open time key program \citep[e.g.,][and references therein]{Muller2009, Lellouch2013, Farkas-Takacs2020} for over 120 TNOs and Centaurs. Nevertheless, stellar occultations provide an alternative means to derive sizes and albedos, potentially with greater precision than thermal data.

The use of stellar occultations is an effective method for directly measuring the sizes and shapes of solar system bodies, often achieving accuracies under 1 km (depending on timing and photometric precision). It also enables probing of their immediate surroundings for features like rings \citep{Braga-Ribas2014, Ortiz2015, Ortiz2017, Morgado2023}, offers the potential to identify binary systems \citep{Leiva2020}, or discover moons \citep[e.g.,][]{Gault2022}. This method can furthermore detect or constrain the presence of atmospheres down to nanobar pressure levels \citep[e.g.,][]{Hubbard1988, Sicardy2003, Oliveira2022}.

Moreover, multi-chord occultation observations yield angular positional measurements of the occulting object with sub-milliarcsecond accuracy, leveraging the high precision of the Gaia reference system \citep[][]{Rommel2020,Ferreira2022,Kaminski2023}. This capability can significantly refine orbital parameters and, consequently, improve predictions for the shadow path of future occultations. Unlike occultations by asteroids, forecasting and successfully observing stellar occultations by TNOs is typically very demanding \citep{Ortiz2020}, primarily due to the minute angular sizes of TNOs coupled with often substantial ephemeris uncertainties.

Trans-Neptunian object (24835) \SM{} was discovered by the 0.9-m Spacewatch telescope at Steward Observatory (Kitt Peak, Arizona) on September 19, 1995 (\href{https://minorplanetcenter.net/mpec/J99/J99L24.html}{MPEC 1999-L24}). This object holds particular significance within the TNO population as one of the few recognized members of a cluster of bodies sharing orbital characteristics highly similar to those of the dwarf planet Haumea. While these Haumea-like bodies have been proposed as a typical collisional family resulting from a disruptive impact \citep{Brown2007}, they do not fully satisfy several criteria for collisional families \citep{Schlichting2009}. Instead, they show closer resemblance to asteroid clusters \citep{Ortiz2019} or minifamilies, which are now understood to form from rotational ejections following a non-disruptive collision \citep{Pravec2018}. A similar formation mechanism was suggested for the Haumea system in \citep{ortiz2012}, which analyzed various rotational fission scenarios. For these reasons, we refer to the set of bodies with orbits related to that of Haumea as a cluster rather than a family. Given that the members would have been formed from the crust of the progenitor, their surfaces should also share characteristics. Indeed, the color and spectrum of \SM{} are highly consistent with those of other cluster members, though its geometric albedo remained an unknown, particularly whether it would align with or significantly differ from Haumea's.

This body, as part of the "TNOs are Cool" project's target list, was observed by the Herschel Space Observatory, but no signal was detected. Consequently, only an upper limit on its size and a lower limit on its albedo could be determined \citep{Vilenius2018} and this lower limit was already unusually high for a TNO, larger than 0.36. This was also the case for other known members of the Haumea orbital cluster  \cite[see][for a review]{Muller2020} . Therefore, \SM{} presented an excellent opportunity for study via the occultation technique to precisely ascertain its size, shape, and geometric albedo.

An extensive observational campaign, involving 50 telescopes, was organized to observe a stellar occultation by \SM{} on February 25, 2024. This effort was prompted by an astrometric update shortly before the event, which indicated a high probability of detection. Given that the occulted star was relatively bright (12.4 mag  in Gaia G filter), making it accessible to numerous instruments, the likelihood of success was considerable. Ultimately, seven positive detections were secured from five different observatories, while 33 sites reported negative results.

In this paper, we present the observations of the stellar occultation by \SM{} and the primary results derived from this event (Sec.~\ref{sec:occultation}). We also incorporate photometric measurements from CCD observations of this target, accumulated over more than a decade from various telescope facilities, to estimate its absolute magnitude, phase slope, and rotational light curve. Finally, we combine all these findings to provide an accurate determination of \SM{}'s geometric albedo (Sec.~\ref{sec:results}). We conclude with an analysis and discussion of these results.

\section{The February 25, 2024 Occultation}\label{sec:occultation}

\subsection{Prediction}\label{sec:prediction}
\begin{figure}[htb]
\includegraphics[width=\columnwidth]
{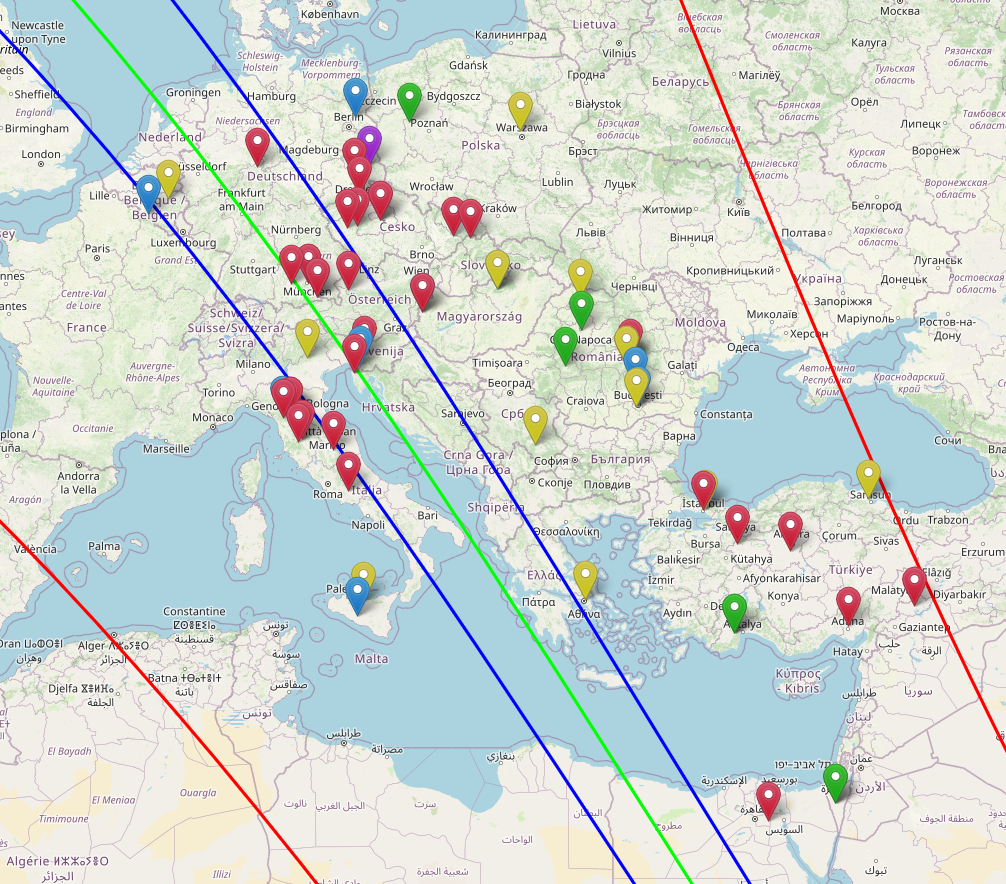}
\caption{This map displays the ground track from our refined prediction, which incorporated astrometry acquired at the 1.5m Sierra Nevada telescope, the 1.2m Calar Alto telescope, and the 2-m Liverpool telescope. The blue lines mark the boundaries of the body's shadow, calculated assuming a spherical shape with a diameter $D = 398$\,km. The central path is indicated by a green line. Red lines illustrate the 3-sigma uncertainties associated with the prediction. The map also shows the observation sites: green markers denote positive detections, red markers indicate negative detections (i.e., 'misses'), blue markers signify planned but unexecuted observations, and yellow markers represent locations with adverse weather. The purple marker indicates technical problems. Map credit: \href{https://www.openstreetmap.org}{OpenStreetMap}.
}\label{fig:prediction+observer}
\end{figure}

Within the framework of the \textsl{Lucky Star} collaboration\footnote{\scriptsize\url{https://lesia.obspm.fr/lucky-star/}}, we generated a prediction for a stellar occultation of a $G = 12.3$ mag star, scheduled for February 25, 2024. This prediction was computed using the Gaia DR3 star catalog in conjunction with NIMA ephemeris~\citep{Desmars2015}. Table~\ref{tab:occdata:2020} summarizes the pertinent occultation parameters and critical details of the occulted star. The specific prediction data presented in Table~\ref{tab:occdata:2020} were sourced from the nominal NIMA (version 9) prediction\footnote{\scriptsize \url{https://lesia.obspm.fr/lucky-star/occ.php?p=126029}}. Some time prior to the occultation date, the prediction was updated and refined through the acquisition of high-precision astrometry. This astrometric data was obtained using the 2-m Liverpool Telescope (LT) at the Roque de Los Muchachos Observatory (ORM) on La Palma, Spain, as well as the 1.5m telescope at Sierra Nevada Observatory (Granada, Spain), the 1.2m telescope at Calar Alto Observatory (Almería, Spain) and the T120 at PHP (Observatoire de Haute Provence, France). This updated information resulted in a eastward shift of the shadow path, directing it into a region with favorable observational prospects (Fig.~\ref{fig:prediction+observer}).

\begin{table}
\centering
\caption{Occultation circumstances and target star parameters for (24835) \SM{} on 25 February 2024.}
\label{tab:occdata:2020}
\begin{tabular}{ll}
\midrule\multicolumn{2}{c}{\textrm{Occultation parameters (NIMAv9)}}\\\midrule
Date and time of CA ($t_0$)			&	Sun. 25 Feb. 2024 \\
									&	18:13:56 UT $\pm 175.6$ s\\
Geocentric shadow velocity				& 	10.42 km/s\\
Magnitude drop						&	8.2~mag\\
Apriori maximum duration 						&   67.6 s\\
Apriori angular size of \SM		    		& 	12 mas \\[1em] 

\midrule\multicolumn{2}{c}{\textrm{Occulted star data (from Gaia DR3)}}\\\midrule
Gaia DR3 source ID					&  224489389287197184	\\
Proper motion (mas/yr)			    		& 	$\mu_\alpha* = +5.8 \pm 0.0$\\
                                    	&	$\mu_\delta  = -7.2 \pm 0.0$\\
Position (ICRS, cat. epoch)			& 	$\alpha$ = 03 40 30.5237\\
									&	$\delta$ = +39 08 06.706\\
Position (ICRS, occ. epoch)			& 	$\alpha$ = 03 40 30.5277\\
									&	$\delta$ = +39 08 06.647\\
Position error (occ. epoch)	        	& 	$\sigma_{\alpha*} = 0.1 \:\textrm{mas}$\\
									&	$\sigma_\delta = 0.1 \:\textrm{mas}$\\
G, RP, BP magnitudes			& 	12.4, 12.0, 12.7 \\
J, H, K magnitudes (from NOMAD) 		& 	11.5, 11.3, 11.3\\

\midrule
\end{tabular}
\begin{minipage}{\columnwidth}
\small
\noindent
The maximum occultation duration (central
line) and the apparent diameter of the TNO were obtained by assuming
a mean geometric TNO albedo, which is far from correct in this case.
V, R, B, J, H, K magnitudes of the target star were taken from the NOMAD catalog \citep{Zacharias2004}.
\end{minipage}
\end{table}

\subsection{Observations}\label{sec:observations}
During the event, meteorological conditions were unfavorable across significant portions of the occultation path. Nevertheless, we successfully obtained seven positive detections from five distinct observatories located in Poland, Romania, Turkey and Israel. Additionally, we recorded instances of near misses both to the South and North of the object (as depicted in Fig.~\ref{fig:prediction+observer}, and detailed in Table~\ref{tab:obs20240225}.

\begin{figure}[tb]
\centering
\includegraphics[width=\columnwidth]{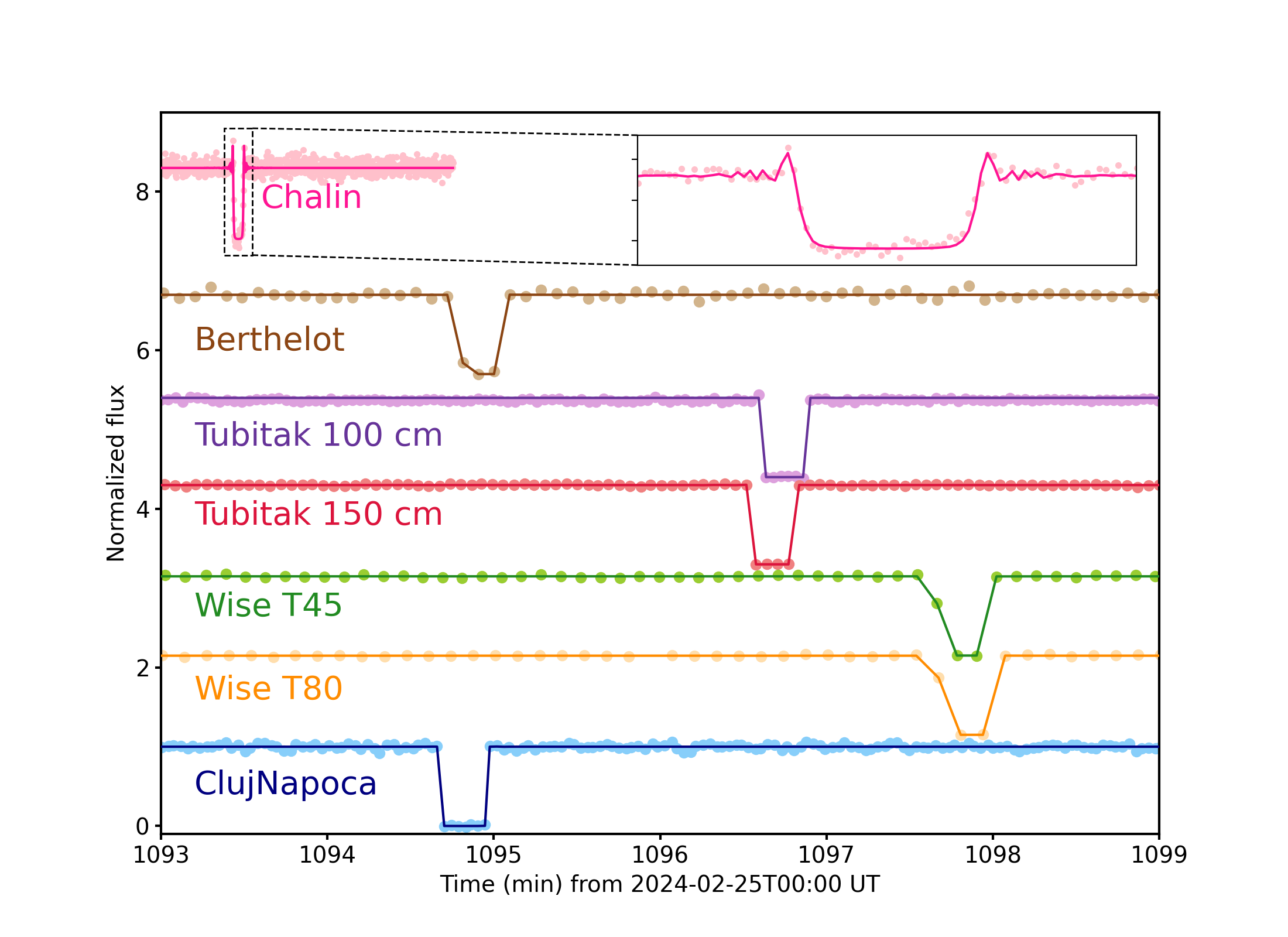}
\caption{Occultation light curves from the various instruments that successfully registered the event. The light curves (flux versus time) are normalized to one, with an arbitrary offset applied for enhanced clarity. Dots represent the observational data, while the lines correspond to the model as described in Section 2. The Chalin light curve  (pink) is presented with an enlarged view in the inset.}\label{fig:olc_all}
\end{figure}

The Occultation Portal~\citep{Kilic2022}\footnote{\scriptsize\url{https://occultationportal.org}} was utilized for observation reporting and data archival. Synthetic aperture photometry was performed on the reduced images to derive the occultation light curves. The seven positive detections are presented in Fig.~\ref{fig:olc_all}.

The star's apparent diameter was determined to be 0.0228 ~mas (in the $V$-band) and 0.0225 mas (in the $B$-band), calculated using formulas published by \cite{Kervella2004}. This corresponds to a projected distance of 0.6\,km at \SM{}'s location, or a duration of 0.06\,s for the shadow's velocity of 10.42\,km/s. The Fresnel scale, $R_F = \sqrt{\lambda \Delta / 2}$, is calculated to be 1.32\,km or 0.134\,s for a wavelength band of $\lambda = 700 \pm 300$\,nm. Given that all positive detections were recorded with exposure times of $\ge 2$\,s, any effects stemming from diffraction or the apparent stellar diameter are negligible, with the specific exception of the Chalin light curve, for which these effects were considered during the determination of ingress and egress times.

The ingress (disappearance) and egress (reappearance) times were extracted from the fitted occultation light curves and subsequently converted into chords on the sky plane. For modeling the light curves and fitting the profile, we employed the SORA Python package \citep{Gomes-Junior2022}. This package also provides capabilities for extracting ingress/egress times using models that account for stellar diameter and diffraction effects where necessary. The extracted timings are listed in Table~\ref{tab:tim2020}.

\begin{table*}[tb]
\caption{Ingress and egress times obtained for the 2024 February 25 occultation. 
}\label{tab:tim2020}
\centering
\begin{tabular}{@{\extracolsep{\fill}}rlllllrc@{}}
\toprule\midrule
\# & Sitename      & Ingress (UT)	& Error   	& Egress (UT) 	& Error   	& Duration  & Chord length \\\midrule
1  &  Astronomical Observatory ClujNapoca    & 18:14:40.87    & 0.82\,s   & 18:14:57.66   & 0.40\,s   & 16.80\,s  &174.9\,km\\
2  & Wise H80      & 18:17:41.091 	& 0.014\,s  & 18:18:00.3 	& 2.4\,s 	& 19.215\,s &  200.1\,km\\
3  & Wise T45.7    & 18:17:40.344 	& 0.035\,s	& 18:17:57.8 	& 2.1\,s 	& 17.456\,s &  181.8\,km\\
4  &  Türkiye National Observatories-150cm & 18:16:32.59 	& 0.95\,s   & 18:16:48.3 	& 1.6\,s 	& 15.7\,s   & 163.6\,km\\ 
5  &  Türkiye National Observatories-100cm & 18:16:36.9 	& 1.2\,s    & 18:16:52.9 	& 1.2\,s 	& 16.0\,s   & 166.7\,km\\ 
6  & Berthelot Observatory     & 18:14:48.534 	& 0.024\,s  & 18:15:02.8 	& 2.3\,s 	& 14.27\,s  & 148.7\,km\\ 
7  & Chalin        & 18:13:26.209	& 0.004\,s  & 18:13:29.618 	& 0.006\,s 	& 3.409\,s  & 35.51\,km\\ 
\midrule
\end{tabular}
\begin{minipage}{\textwidth}
\small
\noindent
Given are the UT times including their respective $1\sigma$-errors, 
the occultation duration in seconds and the corresponding chord length in km.
\end{minipage}
\end{table*}

\subsection{Elliptical fit to the projected shape}
\begin{figure*}
\centering
\includegraphics[width=0.9\columnwidth]{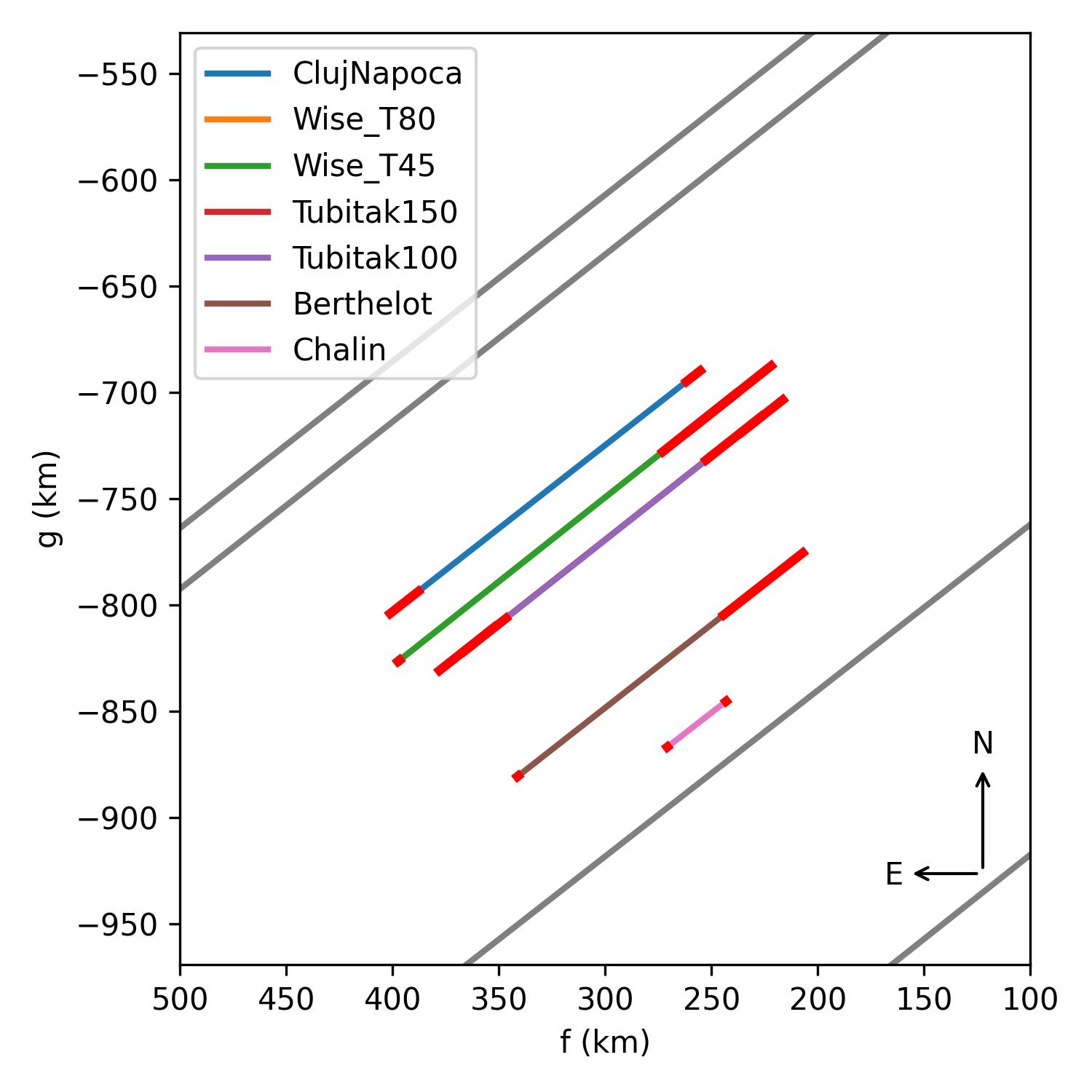}
\includegraphics[width=0.9\columnwidth]{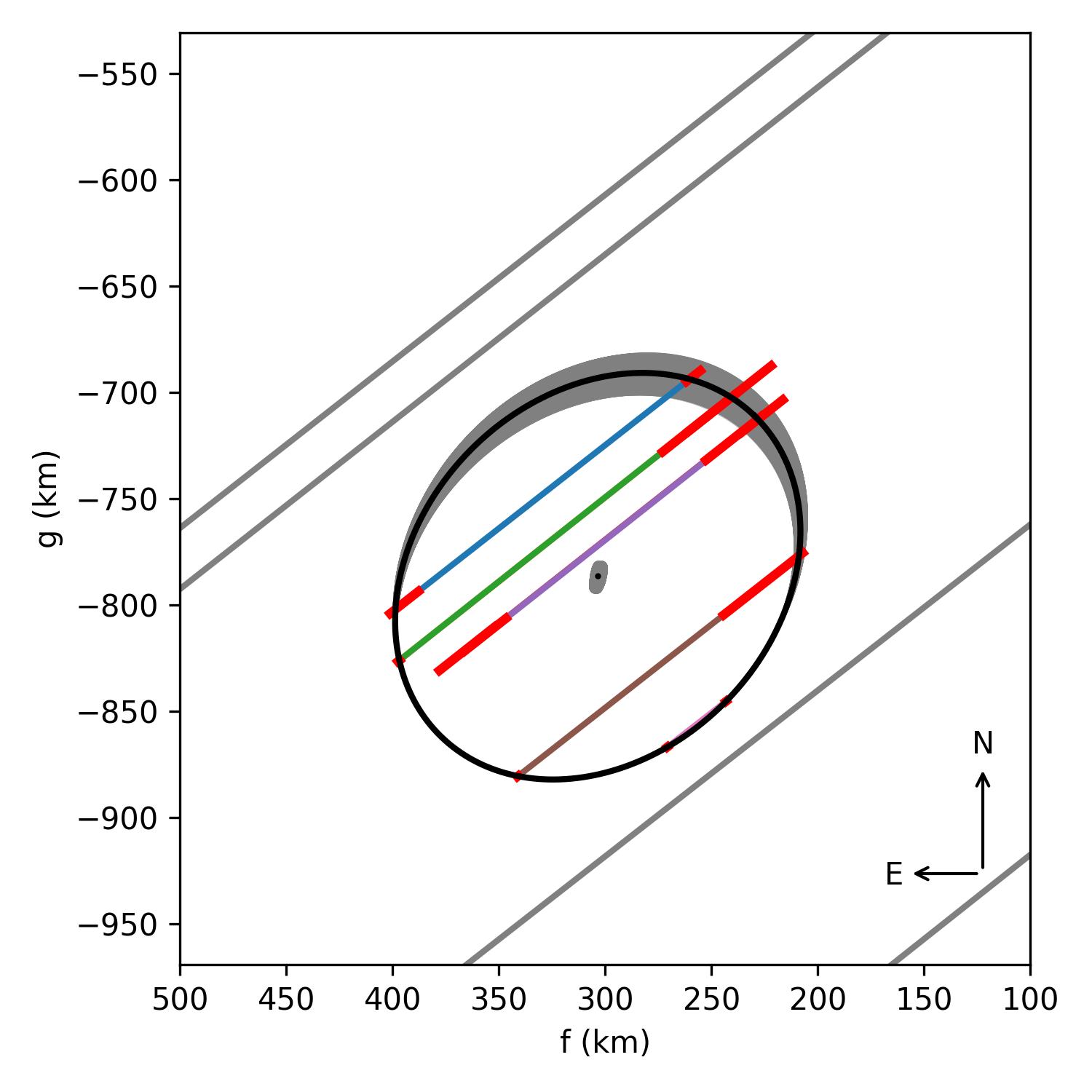}
\caption{Left Panel: The colored segments correspond to positive chords obtained from the sites indicated in the legend. Red segments illustrate the $1\sigma$ uncertainties originating from errors in ingress and egress times.  Note that the large uncertainties come from the fact that the detector was at readout when the reappearance of the star happened. Continuous grey lines denote locations where negative data were obtained. Right panel: This panel displays the elliptical fit to the chords from the 2024 February 25 occultation. This fit characterizes the limb of \SM{} as projected onto the sky plane, defined by the $(f,g)$ axes with origin on NIMA v10 ephemeris, at the moment of the occultation. The two chords obtained using the T100 and RTT150 telescopes at the T\"urkiye National Observatories are graphically indistinguishable in these plots, as are those from the two telescopes at Wise Observatory. 
The grey shaded area represents the $3\sigma$-uncertainty region of the derived ellipse.
}
\label{fig:20240225_profile_fit}
\end{figure*}

Assuming the object's form is either spheroidal or a triaxial ellipsoid, its projection onto the sky plane can be accurately represented as an ellipse. Consequently, we fitted an ellipse to the extremities of the chords, which were derived from the disappearance and appearance times as outlined in Section~\ref{sec:observations}. We also incorporated data from near misses of the occultation as additional constraints (Fig.~\ref{fig:20240225_profile_fit}). The five parameters of the fitted ellipse are: the center of the ellipse $(f,g)$ with respect to the fundamental plane's origin, which is defined by the geocentric star position at the event time and the TNO's ephemeris; the apparent semi-major axis $a'$; the apparent oblateness $\epsilon' = (a'-b')/a'$; and the position angle of the ellipse $\varphi'$\footnote{\scriptsize The (clockwise positive) angle between the `g-positive' direction (i.e. North) and the semi-minor axis $b'$.}. The prime ($'$) notation is used to denote that these parameters correspond to the object's projected ('apparent') profile ellipse, distinguishing them from the axes of a physical body (a triaxial ellipsoid with semi-axes $a,b,c$). These parameters were estimated using a Levenberg-Marquardt optimization algorithm. The goodness of the fit was assessed using the $\chi^2$ per degree of freedom (pdf) value, defined as $\chi^2_\textrm{pdf} = \chi^2 / (N-M)$, where $N=14$ represents the number of data points and $M=5$ is the number of adjustable parameters. An ideal value for this metric is close to one. Our fit yielded $\chi^2_\textrm{pdf} = 1.55$. The $1\sigma$-uncertainties in the retrieved parameters were determined via a grid search in the parameter space, specifically by varying one parameter from its nominal solution while keeping the other parameters constant. Acceptable values were those that produced a $\chi^2$ within the range of $\chi^2_\textrm{min}$ and $\chi^2_\textrm{min} + 1$. The results for our best-fitting instantaneous limb ellipse are summarized in Table~\ref{tab:profile_fit_result}.

\begin{table}[!htb]
\caption{Elliptical fit to the occultation profile.
}\label{tab:profile_fit_result}
\centering
\begin{tabular}{lr}
\toprule\midrule
Center coordinates $(f,g)$					&  	$(303.8 \pm 0.17, -787.8 \pm 0.4)$ km \\
Semi-major axis $a'$						&	$104.3 \pm 0.4$ km \\
Semi-minor axis $b'$						& 	$83.5 \pm 0.5$ km\\
Position angle $\varphi'$					& 	$44.1 \pm 0.4$ deg\\
Oblateness $\epsilon'$						&	$0.200 \pm 0.002$\\
Area-equiv. diameter $D_\textrm{eq,A}$		&	$186.7 \pm 1.8$ km\\
Best-fit $\chi^2_\textrm{pdf}$				&   1.55\\
\midrule
\end{tabular}
\end{table}

\begin{figure}[htb]
\includegraphics[width=\columnwidth]{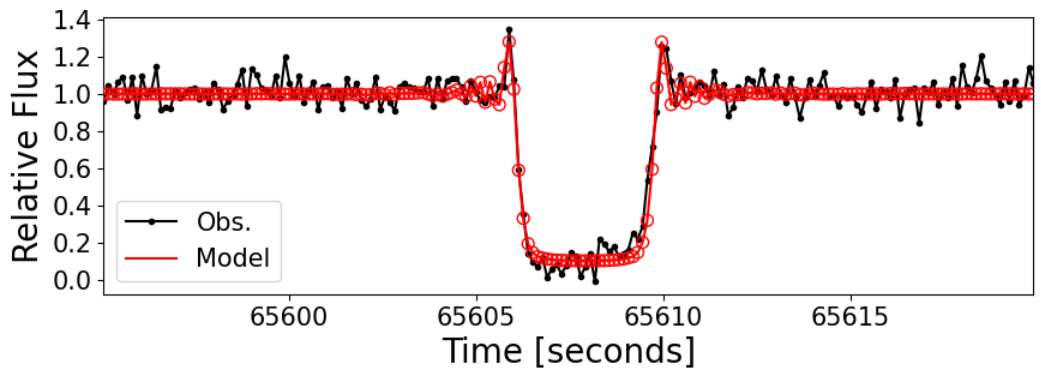}
\caption{Grazing light curve at Chalin (black dots) together with a model fit (red line) as described in the text. The light curve (flux versus time) clearly shows diffraction spikes at both disappearance and reappearance.}\label{fig:Chafit}
\end{figure}

It should be noted that the chord from Chalin is basically  grazing the body as shown in the elliptical fit. In those situations, topographic relief can cause features in the observed occultation light curves. This could be the case here. Hence, we analyzed the curve in some detail here. In fig. ~\ref{fig:Chafit} we show a zoomed version of the light curve where strong Fresnel diffraction effects are observed both at ingress and egress together with a gradual drop to mininum flux and gradual rise to the normal stellar flux. In order to fit those features a smaller velocity of the body with respect to the star is needed than the 10.41 km/s nominal velocity, but this difference can be explained by the inclination between the limb and the occultation path and perhaps the presence of some topography may be playing a small role. There is some structure at the bottom of the light curve but the oscillations are at the noise level and accurate measurements of the height and depth of the potential elevations and depressions cannot be accurately derived. A fit with no topography and a velocity of 4.07 $\pm$ 0.82 km/s is shown in fig. ~\ref{fig:Chafit} to illustrate this point. 

\section{Photometry}\label{sec:photometry}
To comprehensively interpret occultation outcomes concerning an occulting body's three-dimensional shape and size, reliance on a single occultation (which captures a specific rotational phase) is insufficient. Instead, additional stellar occultations at varying rotational phases  i.e. epochs are ideal. Alternatively, leveraging rotational light curves can impose valuable constraints on a TNO's physical model when combined with occultation data, provided that brightness variations are primarily shape-driven rather than albedo-driven. For these reasons, we analyzed existing images of \SM{} from our long-term TNO photometry program and conducted new observations to ascertain its rotational light curve.

Our dataset comprises 649 observations acquired with the 1.5-m telescope at Sierra Nevada Observatory (Spain), the 1.2-m telescope at Calar Alto Observatory (Spain), and the 2-m Liverpool Telescope on La Palma (as discussed in Sec.~\ref{sec:prediction}). Observations with the Liverpool Telescope utilized the IO:O instrument and a Sloan $r'$ filter. At the 1.5-m Sierra Nevada telescope, an Andor iKon-L CCD camera (model DZ936N-BEX2-DD)\footnote{\scriptsize \url{https://www.osn.iaa.csic.es/en/page/ccdt150-and-ccdt90-cameras}} was used, sometimes without filters and other times with Johnson R and V filters. The 1.2-m Calar Alto telescope employed the DLR-MKIII instrument\footnote{\scriptsize \url{https://www.caha.es/es/telescope-1-23m-2/ccd-camera}}, similarly observing without filters or with Johnson R and V filters. Exposure times typically ranged from 300 to 500 seconds. Image reduction and photometric analysis were performed consistently across data from all three telescopes, applying standard bias and flat-field corrections to the raw science images. The observations span from September 2012 to March 2024.

\subsection{Absolute Magnitude}\label{sec:absmag}
\begin{figure}[htb]
\includegraphics[width=\columnwidth]{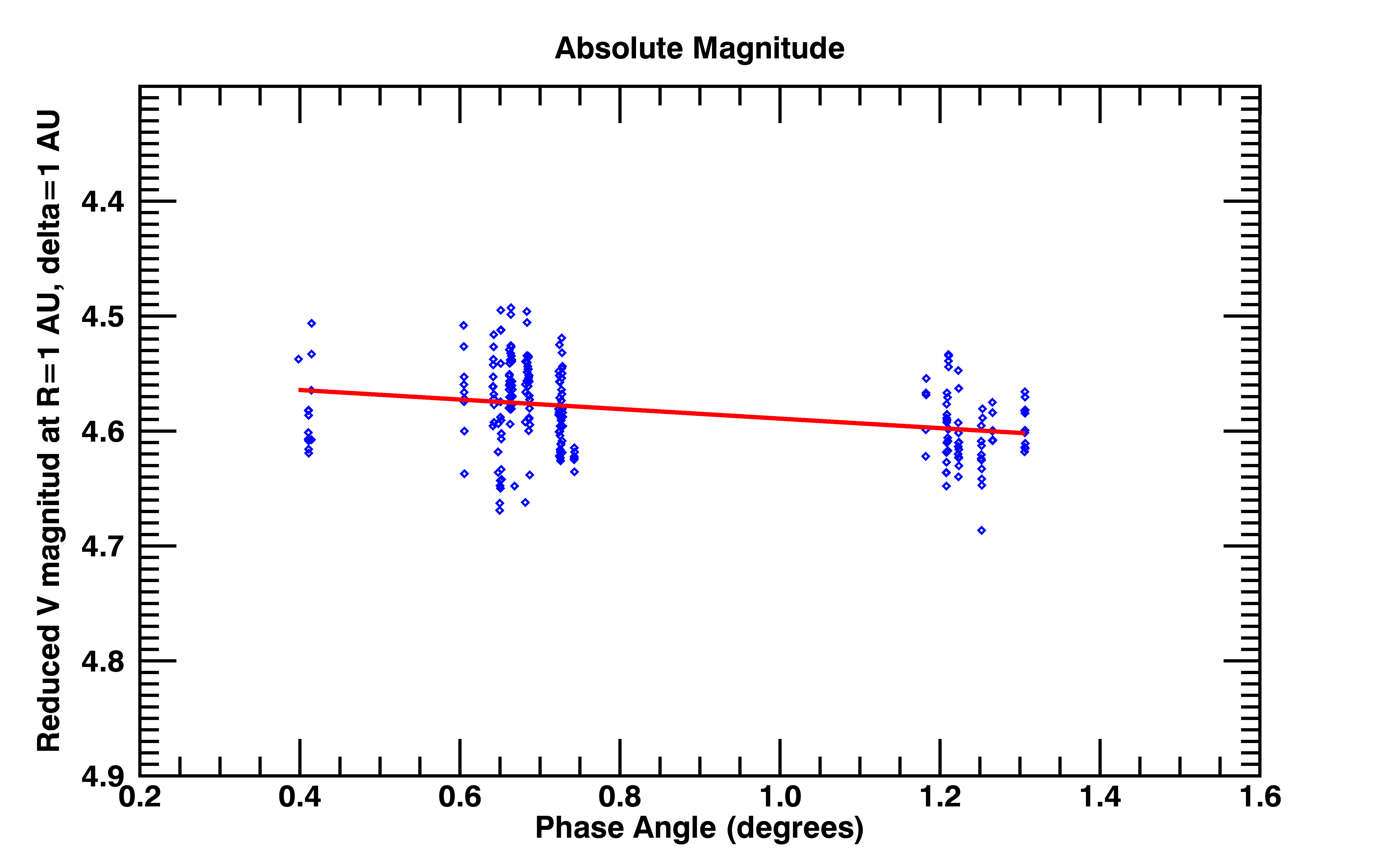}
\caption{Reduced magnitude $m_V(1,1,\alpha)$ plotted against the phase angle $\alpha$. A total of 649 observations, obtained with the 2-m Liverpool telescope, the 1.5-m telescope at Sierra Nevada Observatory, and the 1.2-m telescope at Calar Alto Observatory, were analyzed. This plot was constructed after applying a sigma-clip rejection for outliers and selecting images with a signal-to-noise ratio (SNR) greater than 30.}\label{fig:R11vsPHA}
\end{figure}

From the processed CCD images, we determined magnitudes in the $R$-band and $V$-band using our own algorithms. These algorithms utilize Gaia DR3 field stars to derive photometric transformation equations that incorporate color information~\citep{Morales2022}. For \SM{}'s color, uncorrected for solar color , we adopted a $V-R = 0.37 \pm 0.05$ value, obtained from closely spaced V and R observations at the Calar Alto 1.2-m telescope. This value aligns well with the known color for most members of the Haumea cluster. The $R$-filter dataset was the most extensive. By performing a linear regression of the reduced $R$-magnitude (representing the apparent magnitude the TNO would have at 1\,au from both the Sun and Earth) against the phase angle $\alpha$, we determined the absolute magnitude $m_R(1,1,0)$   ($H_R$ ) and the phase slope $\beta$ (Fig.~\ref{fig:R11vsPHA}). From this fitted trend line, we derived an absolute magnitude $H_R = 4.18 \pm 0.01$ and a slope parameter $\beta = 0.04 \pm 0.02$\,mag/$\degree$. Our observations encompassed a phase angle range of $\alpha = [0.36\degree, 1.54\degree]$. The observed scatter around the trend line may suggest a rotational modulation, albeit with a potential amplitude below 0.1 mag.

\subsection{Rotational Lightcurve}
\begin{figure}[htb]
	\includegraphics[width=\columnwidth]{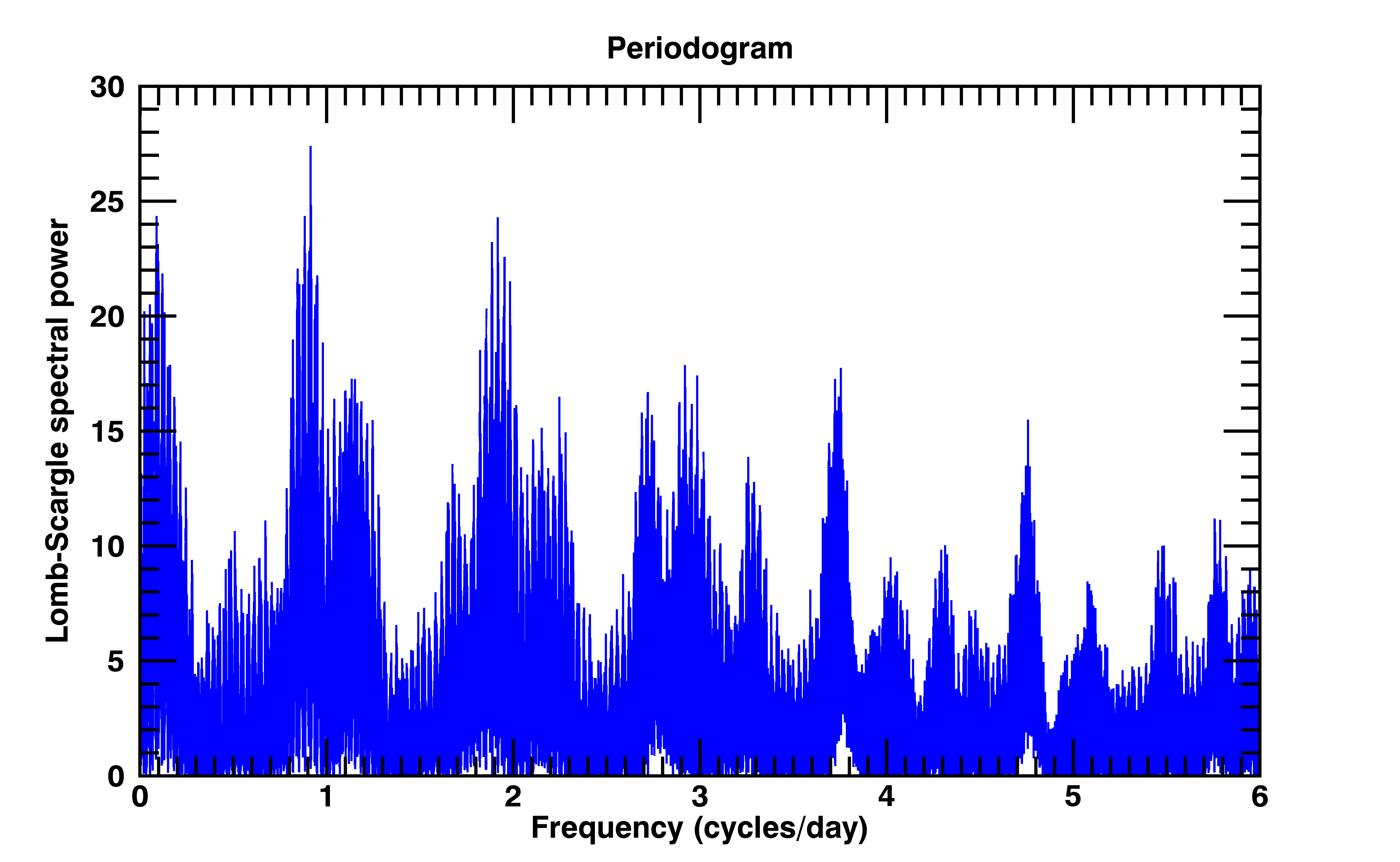}
	\caption{Lomb-Scargle periodogram displaying spectral power versus frequency. 
		The frequency is given in cycles per day to easily identify potential 24-h aliases 
		by their regular spacing. The most prominent peak occurs at 0.91387408 cycles/day, 
		equivalent to 26.26\,h, but other large peaks at $\sim$1.9 cycles/day and $\sim$0.1 cycles/day 
		that appear to be 24-h aliases might actually be the true periodicity. 
		A previous period reported in the literature \citep{Sheppard2003,Thirouin2016} 
		is compatible with the small alias seen at $\sim$5.94 cycles/day in this plot.}
	\label{fig:LSperiodogram}
\end{figure}

After removing the linear phase trend from the photometry (using the O-C residuals of the linear fit discussed in Sec.~\ref{sec:absmag}), and correcting for light-travel time, we proceeded to search for \SM{}'s rotation period using various period-finding techniques. We employed the Lomb-Scargle \citep[L-S;][]{Lomb1976,Scargle1982} algorithm to identify the most probable rotation period from our data. As L-S operates in the frequency domain, the most prominent light curve frequencies (in cycles/day, corresponding to the time scale of the data in days) are displayed in the L-S periodogram (Fig.~\ref{fig:LSperiodogram}). The normalized spectral power reveals a dominant frequency of approximately 0.9139 cycles/day or $P=26.26~h$. Considering that typical small solar system bodies exhibit two maxima and two minima in brightness per rotation period, the most likely rotation frequency would be $f = 0.9139 / 2$ (day$^{-1}$), which corresponds to a rotation period $P = 52.52$~h.
 It is also possible that the TNO could have an oblate shape and the variability could be caused by albedo variations on its surface, giving rise to small variability. In this case the rotation period would be $P=26.26~h$. 

\begin{figure}[htb]
\includegraphics[width=\columnwidth]{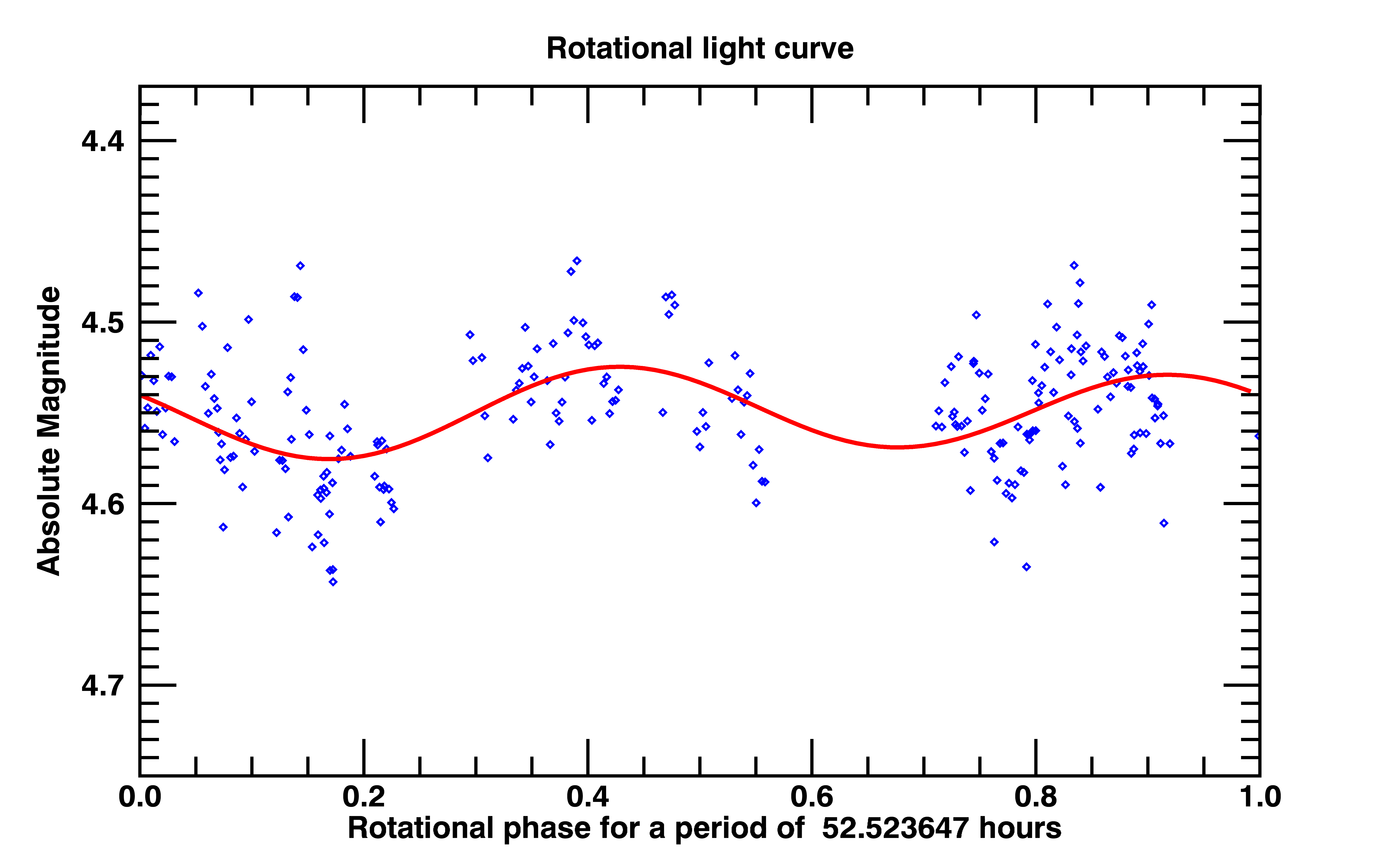}
\caption{Rotational light curve for \SM{} using all photometric data, folded with a period $P=52.52$\,h. It exhibits a double-peaked profile with an amplitude of 0.05\,mag. The time for phase 0.0 was chosen at the moment of mid-occultation.}\label{fig:PhasedRLC}
\end{figure}

We confirmed that folded plots for both periods yielded essentially the same low dispersion relative to a fitted curve,  meaning that we cannot conclude what the cause of the variability is. From the optimal Fourier fit (Fig.~\ref{fig:PhasedRLC}), we determined a peak-to-valley amplitude $\Delta m = 0.05 \pm 0.02$\,mag.

The periodogram shows other high peaks apart from the peak at 0.9139 cycles/day, such as the peaks at $\sim$0.1 cycles/day and $\sim$1.9 cycles/day. They are at frequencies $f_0+k*1.0027 ~or~ f_0-k*1.0027$ where $k$ is an integer and $f_0$ is the main frequency. Thus, they appear to be 24-h aliases of the true frequency, but correctly assessing which one corresponds to the true frequency and which is a 24-h alias is often not possible, especially when the variability is small, as in this case. As noted in \cite{Sheppard2008} where the problem of aliasing is specifically dealt with in TNOs light curves, for low variability objects, very small night-to-night calibration shifts can transfer power to different aliases and identifying the real frequency is problematic. For instance, frequencies of $\sim$0.1 cycles/day and $\sim$1.9 cycles/day would appear possible in our case.
It should be noted that \cite{Sheppard2003} previously proposed a possible rotation period of 8.08\,h for this body, with  a double-peaked light curve and a peak-to-peak variability of 0.04 mag, corresponding to a frequency of 5.94 cycles/day, which would be a 24-h alias at $k=5$ in our interpretation. \cite{Thirouin2016} reached a similar conclusion as that of \cite{Sheppard2003} by combining data from \cite{Sheppard2003} with observations from the 1.5m Sierra Nevada telescope. However, given the significantly extended time span of our observations, which includes campaigns of five consecutive nights in September 2012 and five consecutive nights in December 2013   \citep[the same observations used in][]{Thirouin2016}, we were unable to reproduce that period. Considering that the visibility windows for this target, located at a low southern declination, were limited to only a few hours at the aforementioned observatories, individual runs spanning only a few days may be prone to favoring shorter 24-h aliases of a truly longer period. In contrast, a dataset accumulated over many years can potentially reveal much longer periods. At present, we cannot definitively conclude on the rotation period, but we can affirm that its amplitude is very low.

\section{Results and Discussion}\label{sec:results}
\subsection{Size and Shape}
The instantaneous elliptical limb fit to \SM{}'s projected profile (Fig.~\ref{fig:20240225_profile_fit}) yielded semi-axes $a' = 104.3 \pm 0.3$ km and $b' = 83.5 \pm 0.5$ km, with a position angle $\varphi' = 44.1\degree \pm 0.4\degree$. The goodness of fit, quantified by $\chi^2_\textrm{pdf}$, was $1.55$ (Table~\ref{tab:profile_fit_result}). If the observed light curve intensity variations are attributed to shape effects, this suggests that \SM{} deviates from a perfectly spherical form, although not dramatically, given the modest amplitude of the light-curve. A triaxial ellipsoid with semi-axes $a>b>c$ (where $c$ is the small semi-axis and is in the spin axis direction) offers a suitable approximation for the physical body. In this section, we present potential shapes and sizes of this ellipsoid by combining the occultation observations with our light curve results. According to Maclaurin spheroid theory \citep{Chandrasekhar1969}, a hydrostatically equilibrated body rotating at approximately 52\,h would adopt a Maclaurin spheroid shape with the observed axial ratio if its density were around $\rho \sim 0.03$ g/cm$^3$ and $\rho \sim 0.115$ g/cm$^3$ for a rotation of 26\,h.  Both densities are too low. On the other hand,
realistic Jacobi solutions for triaxial ellipsoids in hydrostatic equilibrium are only feasible for densities that are also too low for bodies of this size range. For all the above, it is clear that \SM{}'s shape is not governed by hydrostatic equilibrium, as expected given that its size is well below the size needed to overcome the strength of the internal material. 

Under the assumption that \SM{}'s shape is triaxial we can make further analyses.The orthogonal projection of a triaxial ellipsoid (with axes $a>b>c$, spinning about $c$) for a given spin state, characterized by the aspect angle $\psi$ (the angle between the rotation axis $c$ and the line of sight) and the rotational phase $\phi$, can be described by \citep[e.g.,][]{Magnusson1986}:
\begin{align}
A  &= b^2c^2 \sin^2\psi \sin^2\phi + a^2c^2 \sin^2\psi\cos^2\phi + a^2b^2\cos^2\psi\\
-B &= a^2(\cos^2\psi\sin^2\phi + \cos^2\phi) + b^2(\cos^2\psi\cos^2\phi + \sin^2\phi) \notag \\&\quad + c^2\sin^2\psi\\
a' &= \left ( \frac{2A}{-B-(B^2-4A)^{1/2}} \right )^{1/2} \label{eq1:magnusson1986} \\
b' &= \left ( \frac{2A}{-B+(B^2-4A)^{1/2}} \right )^{1/2} \label{eq2:magnusson1986},
\end{align}
where $(a',b')$ denote the projected semi-axes, corresponding to the apparent semi-axes of the object's projected cross-section during a stellar occultation. $A$ and $B$ are the coefficients of a second degree equation in q whose solutions are the inverse of the squared semiaxes of the projected ellipse \citep{Drummond1985}.

The rotational light curve amplitude for such an ellipsoid can be calculated using \citep[e.g.,][ p. 426]{Binzel1989}:
\begin{align}\label{eq1:binzel1989}
\Delta m = 2.5 \log \left(\frac{a}{b}\right) - 1.25 \log \left(\frac{a^2\cos^2(\psi) + c^2\sin^2(\psi)}{b^2\cos^2(\psi) + c^2\sin^2(\psi)}\right)
\end{align}

By conducting a grid search across the three body semi-axes $a,b,c$ and the polar aspect angle $\psi$, and applying the rotational phase angle $\phi$ corresponding to the observed occultation time, we can identify the optimal fit to the projected shape derived from the occultation while concurrently fitting the observed rotational light curve amplitude $\Delta m$. The rotational phase at the time of the observed occultation was 0 and the maximum brightness takes place at phase 0.91 (Fig.~\ref{fig:PhasedRLC}) so the rotational phase relative to the maximum of brightness is approximately $0.09$ at the occultation. The observed peak-to-valley amplitude was $\Delta m = 0.05 \pm 0.02$\,mag. We defined the cost function to be minimized as $\chi^2 = (0.05 - \Delta m_c)^ 2 /0.02^2 + (1.25 - a_c'/b_c')^ 2 /0.02^2 + (104.3 - a_c')^2/0.4^2 $, where $\Delta m_c$ is the modeled light-curve amplitude derived from Eq.~\ref{eq1:binzel1989}, and $a_c',b_c'$ are the apparent semi-axes at the rotation phase for the time of the occultation with respect to the maximum obtained from Eqs.~\ref{eq1:magnusson1986}-\ref{eq2:magnusson1986} for each triaxial ellipsoid 'clone' (defined by $a,b,c,\psi$; $\phi=0.09\cdot 2\pi$) generated during the grid search. The parameter space scanned was $c=[60,120]$\,km, $b=[c,150]$ \,km, $a=[b,170]$ \,km, with a grid spacing of 2\,km. The aspect angle $\psi$ was varied from 0 to 90 degrees in $1\degree$ increments. This search yielded a family of potential triaxial ellipsoid solutions and aspect angles. The model that minimizes $\chi^2$ has axes $a = 106\substack{+2\\-2}$\,km, $b = 92\substack{+12\\-12}$\,km, $c = 76\substack{+10\\-10}$\,km, with an aspect angle $\psi = 52\substack{+38\\-28}\degree$. The diameter of an equal-volume sphere for this solution is $D_\textrm{eq} = 181 \pm 12$\,km. The $1\sigma$-uncertainties for the retrieved parameters were obtained by varying one parameter from its nominal solution value (with corresponding $\chi^2=\chi^2_\textrm{min}$) up to $\chi^2=\chi^2_\textrm{min} + 1$, while maintaining other parameters constant. In this way, the triaxial parameters derived from the light curve are consistent with those derived from the occultation. 

\subsection{Absolute magnitude and geometric albedo of \SM{}}
Numerous absolute magnitude data points for \SM{} are available in the literature. \cite{Doressoundiram2002} reported $4.53 \pm 0.02$ at a solar phase angle of $1.3\degree$, assuming a $G$ phase parameter of 0.15. \cite{Romanishin2005} provided ~$H_V=4.54$, also with an assumed $G$ phase parameter of 0.15. \cite{Rabinowitz2008} derived $4.49 \pm 0.03$ with a phase slope parameter of $0.06 \pm 0.03$ mag/$\degree$. \cite{Jewitt2007} observed the object at a phase angle of $0.86\degree$ and applied a slope parameter of $0.04$ mag/$\degree$ in the R band. Using a $V-R = 0.37 \pm 0.07$ color for this body (consistent with \cite{Belskaya2015} for \SM{} and \cite{Verbiscer2022} for Haumea), the absolute magnitude becomes 4.67. \cite{Peixinho2012} quoted $H_R=4.352$, which, with the addition of the $V-R$ value of 0.37, yields $H_V=4.73$. Absolute magnitudes in $g, r, i, z$ bands are also presented by \cite{Ofek2012} in their Table 3. Transforming these $griz$ magnitudes to $UBVRI$ using the equations available online\footnote{\url{https://classic.sdss.org/dr5/algorithms/sdssUBVRITransform.php}} results in an absolute magnitude in $V$ of 4.67.

Our own photometric database, comprising over 400 observations collected spanning more than 10 years, yields an absolute magnitude of $4.55 \pm 0.01$ with a phase slope parameter of $0.04 \pm 0.02$. We found no notable long-term variations within the observation period, suggesting that the spin axis's orientation relative to an Earth observer has remained largely unchanged. Since our observations cover a wide range of phase angles and a substantial time span, effectively smoothing out rotational variability, we maintain confidence in our absolute magnitude determination, which is also consistent with the average of literature values. Given our use of Gaia DR3 stars as calibrators and transformation equations accurate to approximately 0.02 mag, our final estimate is $4.55 \pm 0.03$ mag.

The geometric albedo $p_V$ and diameter $D$ of a small solar system body are linked by the following relation~\citep[e.g.,][]{Russell1916, Harris1998}:
\begin{ceqn}\begin{align}\label{eq:albmag2diam}
D = \frac{D_0}{\sqrt{p_V}} 10 ^{-H_V/5},
\end{align}\end{ceqn}
where $H_V$ is the object's absolute magnitude, $D_0 = 2\,\mathrm{au} \cdot 10^{V_\odot/5}$, and $V_\odot$ is the Sun's apparent V magnitude. Standard values for $V_{\odot}$ include $-26.76$ \citep{Willmer2018} and $-26.74$ \citep{Rieke2008}, which lead to $D_0$ values of 1330.2 km and 1342.6 km, respectively. An older, commonly cited value is $D_0 = 1329$ km.

Applying Eq.~\ref{eq:albmag2diam} with $D_0 = 1330.2$ km and our derived area-equivalent diameter $D = 186.7$ km, and using the absolute magnitude $H_V=4.55$ derived in this work (corrected for the rotational phase at the time of the occultation by 0.02 mag using the rotational light-curve shown in Fig.~\ref{fig:PhasedRLC}), we obtain a geometric albedo of $0.80 \pm 0.04$.

The geometric albedo of Haumea coming from occultation results is reported as $0.51 \pm 0.02$ \citep{Ortiz2017}. \cite{Dunham2019} later refined the 3D shape of Haumea proposed by \cite{Ortiz2017}, suggesting a slightly smaller body in terms of equivalent diameter (both rotational average and volume-equivalent) compared to the original estimate. Based on this reduced size, they proposed a higher albedo of 0.66. However, it is crucial to emphasize that the geometric albedo determined by \cite{Ortiz2017} was derived directly from the accurately measured projected shape at the time of the occultation (without relying on the inferred three-dimensional shape) and incorporated the precise instantaneous absolute magnitude measured concurrently. Therefore, no correction for a smaller overall diameter of Haumea is necessary, and the geometric albedo derived from the occultation remains accurate, with a value of 0.66 being inconsistent with these direct measurements.

Interestingly, the geometric albedo of \SM{} derived in this study is higher than that of Haumea. A similar observation holds for (55636) 2002 TX$_{300}$, another member of the Haumea cluster, for which an occultation of only two chords allowed an albedo determination of $0.88 ^{+0.15}_{-0.06}$ \citep{Elliot2010}. Furthermore, the precise geometric albedo of Haumea's satellite Hi'iaka, obtained from a stellar occultation, is also higher than that of Haumea \citep{Fernandez-Valenzuela2025}. In all three instances, these bodies exhibit a higher geometric albedo than that of Haumea itself.

It is also recognized that the spectral water ice features of these bodies are more pronounced (\cite{Dumas2011}, Pinilla-Alonso private communication), and their phase slope parameters are slightly smaller for \SM{} and 2002 TX$_{300}$ compared to Haumea. These differences suggest that the surface ice has distinct properties on the parent body versus its cluster members and satellites. The underlying reason for this systematic disparity remains unclear. Perhaps some mechanism has caused a darkening of Haumea's surface relative to the surfaces of its cluster members and satellites. While collisional resurfacing of Haumea might tend to replenish its surface with fresh ice from beneath, unexposed to space weathering, it is also plausible that ejecta fallback (which is more significant for Haumea due to its higher escape velocity compared to cluster members) could darken its surface if darker material is excavated during collisional processes. Alternatively, Haumea might experience some form of internal cryovolcanism or another process that alters its surface compared to the rest of the cluster. Cryovolcanism has been hypothesized for TNOs displaying methane ice on their surfaces, such as Eris and Makemake, based on D/H ratios \citep{Grundy2024}, but Haumea lacks methane on its surface. Models concerning the collisional evolution of Haumea's surface \citep{Gil-Hutton2009} indicate that the high abundance of crystalline water ice relative to its amorphous phase might be explained by collisional processes, although no analysis of the effect on geometric albedo was performed. Further insight into the cause of the differing geometric albedo might come from models of multiple-scattered light on these bodies' surfaces, incorporating various particle sizes for ices and slightly different compound mixtures compared to JWST spectra. Additionally, future occultation observations of other members of the Haumea cluster could shed more light on this issue.

\subsection{Astrometry}\label{sec:astrometry}
The ellipse center coordinates presented in Table~\ref{tab:profile_fit_result} constitute two of the five parameters solved for in the ellipse fit. These coordinates represent the observed-computed (O-C) offset between the observed and predicted position (defined by the ephemeris and the star's position). This information is then used to determine the astrometric position of the object. We derived an astrometric position (ICRS) for \SM{} at 2024-02-25~18:13:53.460~UT for a geocentric observer as:
\begin{quote}
\centering
$\alpha$ (hms) = 03 40 30.5344080 $\pm$ 0.1 mas\\
$\delta$ (dms) = +39 08 06.708383 $\pm$ 0.1 mas
\end{quote}
This high-precision astrometry can be employed to determine the TNO's orbit with enhanced accuracy, which will subsequently improve the precision of future occultation predictions.

\section{Conclusions}\label{sec:conclusions}

For the first time, a stellar occultation by the trans-Neptunian object \SM{}, a recognized member of the Haumea orbital cluster, was accurately predicted, subsequently refined, and successfully observed. Making use of seven occultation chords gathered from five distinct observing sites, we determined the object's instantaneous projected shape and size by fitting an elliptical profile with semiaxes dimensions of $(104.3 \pm 0.3) \times (83.5 \pm 0.5)$~km. The computed area-equivalent diameter at the time of the occultation is $D_\mathrm{eq,A} = 186.7 \pm 1.8$~km.

Additionally, our study yielded an absolute magnitude of $H_V = 4.55 \pm 0.03$, a $V-R$ color of $0.37 \pm 0.05$, and a phase slope of $0.04 \pm 0.02$\,mag/$\degree$. These values are consistent with prior research. Furthermore, we attempted to ascertain this TNO's rotation period from our extensive photometric observation campaigns. However, a definitive rotation period could not be established with certainty. Our preferred rotation period $~P~$ is ~$52.52 \pm 0.02$~hours  or ~$26.26 \pm 0.01$~hours. The light curve is either double-peaked or single-peaked, a distinction that could not be conclusively made, exhibiting a peak-to-valley amplitude $\Delta m = 0.05 \pm 0.02$\,mag.

By integrating the occultation data with the rotational light curve results, we derived some constraints on \SM{}'s 3D size and shape. Nevertheless, these constraints remain weak due to significant uncertainties in the aspect angle and rotational properties.

From the derived area-equivalent diameter ~$D_\mathrm{eq,A} = 186.7 \pm 1.8$~km, the aforementioned absolute magnitude, and applying a 0.02 mag correction for the rotational phase at the time of the occultation, we calculated a geometric albedo of $p_V = 0.80 \pm 0.04$. This value is notably higher than the geometric albedo reported for Haumea. This trend also appears to extend to other members of the Haumea orbital cluster and its satellites, suggesting systematic differences in the ice composition or properties covering their surfaces, compared to Haumea.

Finally, we successfully derived an occultation-based astrometric position (ICRS) for (24835) \SM{} (Sec.~\ref{sec:astrometry}).

\begin{acknowledgements}
This work was supported by multiple funding agencies and institutions. It was partly funded by the Spanish projects PID2020-112789GB-I00 (AEI) and Proyecto de Excelencia de la Junta de Andalucía PY20-01309. J.L.O., P.S.-S., N.M., A.A.C and R.D. acknowledge financial support from the Severo Ochoa grant CEX2021-001131-S (MCIN/AEI/10.13039/501100011033). P.S.-S. also acknowledges support from the Spanish I+D+i project PID2022-139555NB-I00 (TNO-JWST) funded by MCIN/AEI/10.13039/501100011033. AAC acknowledges financial support from the project PID2023-153123NB-I00 funded by MCIN/AEI. Gy.M.Sz. acknowledges the SNN-147362, GINOP-2.3.2-15-2016-00003, and K-138962 grants of the Hungarian Research, Development and Innovation Office (NKFIH). Z.G. acknowledges the PRODEX Experiment Agreement No. 4000137122 between ELTE Eötvös Loránd University and ESA, the VEGA grant No. 2/0031/22 of the Slovak Academy of Sciences, the Slovak Research and Development Agency contract No. APVV-20-0148, and support from the city of Szombathely. J.I.B.C. acknowledges CNPq grants 305917/2019-6 and 306691/2022-1, and FAPERJ grant 201.681/2019. F.B.R. acknowledges CNPq grant 316604/2023-2. This study was financed in part by CAPES (Finance Code 001) and the National Institute of Science and Technology of the e-Universe project (INCT do e-Universo, CNPq grant 465376/2014-2). A. Takey and A.M. Abdelaziz acknowledge financial support from the Egyptian Science, Technology \& Innovation Funding Authority (STDF) under grant 48102. M.A. acknowledges grants 427700/2018-3, 310683/2017-3, and 473002/2013-2. The work of A.S. and D.A.N. was supported by a grant of the Ministry of Research, Innovation and Digitalization (CCCDI – UEFISCDI, project PN-IV-P6-6.3-SOL-2024-2-0220, within PNCDI IV). D.I. acknowledges funding provided by the University of Belgrade – Faculty of Mathematics through grant 451-03-136/2025-03/200104 from the Ministry of Science, Technological Development and Innovation of the Republic of Serbia. V.N. acknowledges support from the Bando Ricerca Fondamentale INAF 2023 Data Analysis Grant: "Characterization of transiting exoplanets by exploiting the unique synergy between TASTE and TESS". Operation of the University of Haifa's H80 telescope at the Wise Observatory is partly supported by ISF grant 3200/20. .{I}ST60 and .{I}ST40 are observational facilities of Istanbul University Observatory, funded by the Istanbul University Scientific Research Projects Coordination Unit (projects BAP-3685 and FBG-2017-23943) and by the Presidency of Strategy and Budget of the Republic of Türkiye (project 2016K12137). This work is partly based on observations collected at the Centro Astronómico Hispano en Andalucía (CAHA), Observatorio de Sierra Nevada (IAA-CSIC), and the Liverpool Telescope at the Roque de los Muchachos Observatory (IAC). This research has made use of data from the European Space Agency (ESA) mission Gaia (\url{https://www.cosmos.esa.int/gaia}), processed by the Gaia Data Processing and Analysis Consortium (\url{https://www.cosmos.esa.int/web/gaia/dpac/consortium}), with funding provided by institutions participating in the Gaia Multilateral Agreement. The authors thank Peter C. Slansky and M. Krahn for their observational contribution, and gratefully acknowledge all observers who attempted to observe this occultation event but are not explicitly mentioned in Table~\ref{tab:obs20240225}. They also thank their collaborators at the University of Athens Observatory for utilizing the robotic Cassegrain reflector (see \cite{Gazeas2016} for details).

\end{acknowledgements}


\bibliographystyle{aa}
\bibliography{references}


\begin{appendix}

\onecolumn
\section{Observation Details}

\begin{small}
\begin{longtable}{@{}r
  l        
  l        
  l        
  l        
  l        
  @{\hspace{20pt}}
  l        
  l        
  l        
  l        
  l        
  @{\hspace{20pt}}
  l        
@{}}
\caption{Observation details for the 2024 February 25 occultation by \SM{}.}\label{tab:obs20240225}\\
\toprule\midrule
\# & Site name & Country & Latitude (dms) & Telescope & Method & Occultation \\
  & Observer(s) &  & Longitude (dms) & Camera & ExpTime & DeadTime \\
  &  &  & Elevation (m) & Filter & TimeSrc &  \\
\midrule
\endfirsthead

\caption[]{Observation details for the 2024 February 25 occultation by \SM{} (continued).}\\
\toprule\midrule
\# & Site name & Country & Latitude (dms) & Telescope & Method & Occultation \\
  & Observer(s) &  & Longitude (dms) & Camera & ExpTime & DeadTime \\
  &  &  & Elevation (m) & Filter & TimeSrc &  \\
\midrule
\endhead

\midrule
\endfoot

\bottomrule
\multicolumn{11}{p{\textwidth}}{\footnotesize
Site latitude, longitude (format dms) and elevation (AMSL in m) are given in the WGS84 datum.
\textbf{Telescope}: Tx refers to the telescope aperture in cm.
\textbf{Method} is the recording method: \textbf{IMG} means digital (CCD, CMOS) sequential imaging, while \textbf{VID} means analogue video recording.
\textbf{TimeSrc} refers to the used timing source and method: \textbf{GPS}: 1-PPS (one pulse per second) driven video-time-insertion (VID) or camera-internal GPS timestamps (IMG). \textbf{NTP}: Network Time Protocol computer system clock synchronization. \textbf{CamGPS}: Camera synchronized directly to a GPS signal for timestamping accuracy. \textbf{ComGPS}: Computer synchronized to a GPS signal. \textbf{ComNTP}: Computer synchronized using the Network Time Protocol. \textbf{TimeBox}: A TimeBox is used for precise timestamping during observations. \textbf{IOTA-VTI}: A hardware device developed by IOTA (International Occultation Timing Association) for video time insertion. \textbf{Other}: Any other synchronization method not listed above.
Observation can be either positive (occultation detected/recorded) or negative (occultation not detected).
\textbf{ExpTime} represents the exposure time in seconds, while \textbf{DeadTime} refers to the interval between subsequent images in seconds.  The sampling cadence is the sum of ExpTime and DeadTime.
}\\
\endlastfoot

    1 & Astronomical Observatory Cluj-Napoca, & Romania & 46$^\circ$ 42$^\prime$ 37.557$^{\prime\prime}$ N& 61.0 cm& IMG & Positive \\
    & Feleacu Station & & 23$^\circ$ 35$^\prime$ 35.6479$^{\prime\prime}$ E & SBIG STT1603ME & 1\,s & 1.6\,s\\
    & \multicolumn{1}{l}{\parbox[t]{5.5cm}{\raggedright\em Vlad Turcu, Dan Moldovan}} & & 783.40 & Empty & CamGPS &\\\midrule
    2 & Berthelot Observatory & Romania & 45$^\circ$ 36$^\prime$ 59.2515$^{\prime\prime}$ N& 50.0 cm& IMG & Positive \\
    &  & & 22$^\circ$ 53$^\prime$ 19.6894$^{\prime\prime}$ E & SBIG STXL-6303 & 1\,s & 4.7\,s\\
    & \multicolumn{1}{l}{\parbox[t]{5.5cm}{\raggedright\em Adrian Sonka, Elisabeta Petrescu, Alin Nedelcu}} & & 386.36 & Clear & ComNTP &\\\midrule
    3 & T\"urkiye National Observatories & T\"urkiye & 36$^\circ$ 49$^\prime$ 32.1948$^{\prime\prime}$ N& 150.0 cm& IMG & Positive \\
    & (RTT150) & & 30$^\circ$ 20$^\prime$ 07.25715$^{\prime\prime}$ E & Andor iKon-L 936 & 0.1\,s & 3.78\,s\\
    & \multicolumn{1}{l}{\parbox[t]{5.5cm}{\raggedright\em I. Akoz, Y. Kilic, O. Erece, K. Uluc}} & & 2458.59 & Clear & CamGPS &\\\midrule
    4 & T\"urkiye National Observatories & T\"urkiye & 36$^\circ$ 49$^\prime$ 17.0672$^{\prime\prime}$ N& 100.0 cm& IMG & Positive \\
    & (T100) & & 30$^\circ$ 20$^\prime$ 07.98456$^{\prime\prime}$ E & SI 1100 & 0.3\,s & 2.36\,s\\
    & \multicolumn{1}{l}{\parbox[t]{5.5cm}{\raggedright\em Y. Kilic, O. Erece, C. Nehir, K. Uluc}} & & 2538.72 & Clear & CamGPS &\\\midrule
    5 & Wise & Israel & 30$^\circ$ 35$^\prime$ 48.5862$^{\prime\prime}$ N& 80 cm& IMG & Positive \\
    &  & & 34$^\circ$ 45$^\prime$ 44.1366$^{\prime\prime}$ E & STL6303 & 3.0\,s & 5.0\,s\\
    & \multicolumn{1}{l}{\parbox[t]{5.5cm}{\raggedright\em Shai Kaspi}} & & 862.27 & Empty & ComNTP &\\\midrule
    6 & Wise & Israel & 30$^\circ$ 35$^\prime$ 48.5862$^{\prime\prime}$ N& 45.7 cm& IMG & Positive \\
    &  & & 34$^\circ$ 45$^\prime$ 44.1366$^{\prime\prime}$ E & QSI683 & 3.0\,s & 4.1\,s\\
    & \multicolumn{1}{l}{\parbox[t]{5.5cm}{\raggedright\em Shai Kaspi}} & & 862.27 & Empty & ComNTP &\\\midrule
    7 & Chalin & Poland & 52$^\circ$ 36$^\prime$ 14.027$^{\prime\prime}$ N& 35.0 cm& IMG & Positive \\
    &  & & 16$^\circ$ 02$^\prime$ 33.4403$^{\prime\prime}$ E & Andor Zyla 5.5 & 0.1\,s & 0.027\,s\\
    & \multicolumn{1}{l}{\parbox[t]{5.5cm}{\raggedright\em A. Marciniak}} & & 65.00 & Empty & CamGPS &\\\midrule
    8 & Wendelstein Observatory & Germany & 47$^\circ$ 42$^\prime$ 14.5368$^{\prime\prime}$ N& 210.0 cm& IMG & Negative \\
    &  & & 12$^\circ$ 00$^\prime$ 47.88$^{\prime\prime}$ E & WNIR & -\ & 1.4336\,s\\
    & \multicolumn{1}{l}{\parbox[t]{5.5cm}{\raggedright\em Michael Schmidt}} & & 1931.00 & J & ComNTP &\\\midrule
    9 & KAO & Egypt & 29$^\circ$ 56$^\prime$ 2.4$^{\prime\prime}$ N& 188.0 cm& IMG & Negative \\
    &  & & 31$^\circ$ 49$^\prime$ 37.2$^{\prime\prime}$ E & E2V 42-40 2k & 4\,s & 8.3\,s\\
    & \multicolumn{1}{l}{\parbox[t]{5.5cm}{\raggedright\em Ali Takey, A. E. Abdelaziz}} & & 476.00 & UBVRI/SDSS-ugriz & CamGPS &\\\midrule
    10 & ELTE GAO MKK & Hungary & 47$^\circ$ 15$^\prime$ 28.5471$^{\prime\prime}$ N& 80.0 cm& IMG & Negative \\
    &  & & 16$^\circ$ 36$^\prime$ 12.1097$^{\prime\prime}$ E & Photometrics Prime 95B & 0.4\,s & 0.0001\,s\\
    & \multicolumn{1}{l}{\parbox[t]{5.5cm}{\raggedright\em Gy. M. Szabó, J. Kovács, Z. Garai}} & & 209.00 & Empty & ComGPS &\\\midrule
    11 & Bavarian Public Observatory & Germany & 48$^\circ$ 07$^\prime$ 18.9984$^{\prime\prime}$ N& 80.0 cm& IMG & Negative \\
    & Munich & & 11$^\circ$ 36$^\prime$ 25.9992$^{\prime\prime}$ E & ASI1600 & 0.04\,s & 0.0\,s\\
    & \multicolumn{1}{l}{\parbox[t]{5.5cm}{\raggedright\em Bernd Gaehrken}} & & 500.00 & Empty & Other &\\\midrule
    12 & Ondrejov & Czechia & 49$^\circ$ 54$^\prime$ 38.016$^{\prime\prime}$ N& 65.0 cm& IMG & Negative \\
    &  & & 14$^\circ$ 47$^\prime$ 1.104$^{\prime\prime}$ E & G2CCD-3200 & 3.0\,s & 1.2\,s\\
    & \multicolumn{1}{l}{\parbox[t]{5.5cm}{\raggedright\em K. Hornoch}} & & 528.00 & Empty & ComNTP &\\\midrule
    13 & Observatory Teplice & Czechia & 50$^\circ$ 38$^\prime$ 17.9556$^{\prime\prime}$ N& 60.0 cm& IMG & Negative \\
    &  & & 13$^\circ$ 50$^\prime$ 48.3$^{\prime\prime}$ E & FLI Kepler FL4040 & 0.1\,s & 0.2\,s\\
    & \multicolumn{1}{l}{\parbox[t]{5.5cm}{\raggedright\em Tomas Janík}} & & 277.94 & Empty & CamGPS &\\\midrule
    14 & San Marcello Pistoiese & Italy & 44$^\circ$ 03$^\prime$ 46.9296$^{\prime\prime}$ N& 60.0 cm& IMG & Negative \\
    &  & & 10$^\circ$ 48$^\prime$ 15.12$^{\prime\prime}$ E & Apogee & 1\,s & 1.0\,s\\
    & \multicolumn{1}{l}{\parbox[t]{5.5cm}{\raggedright\em P. Bacci, M. Maestripieri}} & & 965.41 & Empty & ComNTP &\\\midrule
    15 & ?rni Vrh Observatory & Slovenia & 45$^\circ$ 56$^\prime$ 45.0688$^{\prime\prime}$ N& 60.0 cm& IMG & Negative \\
    &  & & 14$^\circ$ 04$^\prime$ 16.6242$^{\prime\prime}$ E & ZWO ASI6200MM Pro & 0.5\,s &  0.5\,s\\
    & \multicolumn{1}{l}{\parbox[t]{5.5cm}{\raggedright\em J. Skvarc}} & & 713.03 & Empty & ComNTP &\\\midrule
    16 & Montarrenti Observatory & Italy & 43$^\circ$ 13$^\prime$ 57$^{\prime\prime}$ N& 53.0 cm& IMG & Negative \\
    &  & & 11$^\circ$ 11$^\prime$ 2.04$^{\prime\prime}$ E & Apogee Alta U47 & 1\,s & 1.3\,s\\
    & \multicolumn{1}{l}{\parbox[t]{5.5cm}{\raggedright\em S. Leonini, M. Conti, P. Rosi, L.M. Tinjaca Ramirez, L. Bellizzi}} & & 347.00 & Clear & ComNTP &\\
    \\ \pagebreak
    17 & Monte Agliale Observatory & Italy & 43$^\circ$ 59$^\prime$ 43.008$^{\prime\prime}$ N& 50.0 cm& IMG & Negative \\
    &  & & 10$^\circ$ 30$^\prime$ 53.496$^{\prime\prime}$ E & SBIG ST9 & 3.0\,s & 2.0\,s\\
    & \multicolumn{1}{l}{\parbox[t]{5.5cm}{\raggedright\em F. Ciabattari}} & & 760.00 & Empty & ComNTP &\\\midrule
    18 & Harpoint Observatory & Austria & 47$^\circ$ 54$^\prime$ 33.12$^{\prime\prime}$ N& 50.0 cm& IMG & Negative \\
    &  & & 13$^\circ$ 21$^\prime$ 6.12$^{\prime\prime}$ E & ZWO ASI 6200MM & 0.1\,s & 0.4\,s\\
    & \multicolumn{1}{l}{\parbox[t]{5.5cm}{\raggedright\em R. Schaefer}} & & 700.00 & Empty & ComGPS &\\\midrule
    19 & Sternwarte Radebeul & Germany & 51$^\circ$ 6$^\prime$ 58.6728$^{\prime\prime}$ N& 43.2 cm& IMG & Negative \\
    &  & & 13$^\circ$ 37$^\prime$ 19.92$^{\prime\prime}$ E & QHY600 & 0.2\,s & 1.0\,s\\
    & \multicolumn{1}{l}{\parbox[t]{5.5cm}{\raggedright\em Martin Fiedler}} & & 185.00 & Astronomik L2 & ComNTP &\\\midrule
    20 & Wendelstein Observatory & Germany & 47$^\circ$ 42$^\prime$ 14.5368$^{\prime\prime}$ N& 43.0 cm& IMG & Negative \\
    &  & & 12$^\circ$ 0$^\prime$ 47.88$^{\prime\prime}$ E & QHY 600 M Pro & -\ & 0.4\,s\\
    & \multicolumn{1}{l}{\parbox[t]{5.5cm}{\raggedright\em Michael Schmidt}} & & 1931.00 & g' & ComNTP &\\\midrule
    21 & Starhopper Observatory & Romania & 45$^\circ$ 51$^\prime$ 56.0016$^{\prime\prime}$ N& 40.6 cm& IMG & Negative \\
    &  & & 25$^\circ$ 46$^\prime$ 8.0004$^{\prime\prime}$ E & ASI6200MMPRO & 5\,s & 0.5\,s\\
    & \multicolumn{1}{l}{\parbox[t]{5.5cm}{\raggedright\em Felician Ursache}} & & 588.00 & Empty & ComNTP &\\\midrule
    22 & ?stanbul University Observatory & T\"urkiye & 41$^\circ$ 0$^\prime$ 42.2964$^{\prime\prime}$ N& 40.0 cm& IMG & Negative \\
    & Application and Research Center (?ST40) & & 28$^\circ$ 57$^\prime$ 56.5848$^{\prime\prime}$ E & Moravian G2 8300 & 2.0\,s & 0\,s\\
    & \multicolumn{1}{l}{\parbox[t]{5.5cm}{\raggedright\em S. Fi?ek, S. Alis, F. K. Yelkenci}} & & 60.00 & Clear & ComNTP &\\\midrule
    23 & Karrenkneul & Hessen & 51$^\circ$ 23$^\prime$ 36.8938$^{\prime\prime}$ N& 40.0 cm& IMG & Negative \\
    &  & & 9$^\circ$ 21$^\prime$ 38.6691$^{\prime\prime}$ E & ZWO ASI 6200 pro MM & 0.2\,s & 0.1\,s\\
    & \multicolumn{1}{l}{\parbox[t]{5.5cm}{\raggedright\em Martin Krahn}} & & 266.49 & Luminance & ComNTP &\\\midrule
    24 & DROT & Slovakia & 49$^\circ$ 24$^\prime$ 15.2341$^{\prime\prime}$ N& 40.0 cm& IMG & Negative \\
    &  & & 18$^\circ$ 42$^\prime$ 9.33781$^{\prime\prime}$ E & QHY5III-290M & 0.05\,s & 0.01\,s\\
    & \multicolumn{1}{l}{\parbox[t]{5.5cm}{\raggedright\em Peter Delincak}} & & 635.00 & Clear & ComNTP &\\\midrule
    25 & Ondrejov Cosmic Lab roof & Czechia & 49$^\circ$ 54$^\prime$ 36.2016$^{\prime\prime}$ N& 35.6 cm& IMG & Negative \\
    &  & & 14$^\circ$ 46$^\prime$ 47.7012$^{\prime\prime}$ E & DVTI+CAM 430 & 0.100\,s & 0.0001\,s\\
    & \multicolumn{1}{l}{\parbox[t]{5.5cm}{\raggedright\em J. Mánek}} & & 524.00 & Empty & CamGPS &\\\midrule
    26 & Roof Observatory Kaufering & Germany & 48$^\circ$ 5$^\prime$ 22.7793$^{\prime\prime}$ N& 35.0 cm& IMG & Negative \\
    &  & & 10$^\circ$ 50$^\prime$ 57.9958$^{\prime\prime}$ E & QHY174M-GPS & 0.1\,s & <0.001\,s\\
    & \multicolumn{1}{l}{\parbox[t]{5.5cm}{\raggedright\em Gregor Krannich}} & & 596.00 & Empty & CamGPS &\\\midrule
    27 & Strašice & Czechia & 49$^\circ$ 44$^\prime$ 35.7118$^{\prime\prime}$ N& 30.3 cm& IMG & Negative \\
    &  & & 13$^\circ$ 44$^\prime$ 56.3093$^{\prime\prime}$ E & QHY174M-GPS & 0.15\,s & 0.0\,s\\
    & \multicolumn{1}{l}{\parbox[t]{5.5cm}{\raggedright\em Ji?í Kubánek}} & & 536.61 & Empty & CamGPS &\\\midrule
    28 & Astronomical Observatory, University & Italia & 43$^\circ$ 18$^\prime$ 45$^{\prime\prime}$ N& 30.0 cm& IMG & Negative \\
    & of Siena (Italy) & & 11$^\circ$ 20$^\prime$ 12.12$^{\prime\prime}$ E & Sbig STL-6303 & 1.0\,s & 3.0\,s\\
    & \multicolumn{1}{l}{\parbox[t]{5.5cm}{\raggedright\em A. Marchini}} & & 290.00 & Clear & ComNTP &\\\midrule
    29 & Plzen-Valcha & Czechia & 49$^\circ$ 42$^\prime$ 26.4403$^{\prime\prime}$ N& 25.4 cm& IMG & Negative \\
    &  & & 13$^\circ$ 19$^\prime$ 55.4668$^{\prime\prime}$ E & QHY174M-GPS & 0.100\,s & 0.0\,s\\
    & \multicolumn{1}{l}{\parbox[t]{5.5cm}{\raggedright\em M.Rottenborn }} & & 325.38 & none & CamGPS &\\\midrule
    30 & Observatory Valasske Mezirici & Czechia & 49$^\circ$ 27$^\prime$ 49.2984$^{\prime\prime}$ N& 25.4 cm& IMG & Negative \\
    &  & & 17$^\circ$ 58$^\prime$ 25.2984$^{\prime\prime}$ E & QHY174M-GPS & 0.500\,s & 0.1\,s\\
    & \multicolumn{1}{l}{\parbox[t]{5.5cm}{\raggedright\em Petr Zeleny}} & & 338.00 & Empty & CamGPS &\\\midrule
    31 & Volkssternwarte Muenchen & Germany & 48$^\circ$ 7$^\prime$ 19.0733$^{\prime\prime}$ N& 25.0 cm& IMG & Negative \\
    &  & & 11$^\circ$ 36$^\prime$ 25.5974$^{\prime\prime}$ E & Sony Alpha 7S & 0.080\,s & 0.0\,s\\
    & \multicolumn{1}{l}{\parbox[t]{5.5cm}{\raggedright\em Peter C. Slansky}} & & 565.00 & None & CamGPS &\\\midrule
    32 & Frosinone & Italy & 41$^\circ$ 39$^\prime$ 11.6387$^{\prime\prime}$ N& 23.5 cm& IMG & Negative \\
    &  & & 13$^\circ$ 20$^\prime$ 15.7991$^{\prime\prime}$ E & QHY294M Pro & 0.394\,s & 0.0\,s\\
    & \multicolumn{1}{l}{\parbox[t]{5.5cm}{\raggedright\em D. Pica}} & & 198.98 & Clear & ComNTP &\\\midrule
    33 & Brtonigla & Croatia & 45$^\circ$ 23$^\prime$ 4.99999$^{\prime\prime}$ N& 20.0 cm& IMG & Negative \\
    &  & & 13$^\circ$ 37$^\prime$ 47$^{\prime\prime}$ E & ZWO ASI178 MM & 1.0\,s & 0.0\,s\\
    & \multicolumn{1}{l}{\parbox[t]{5.5cm}{\raggedright\em H. Mikuz, A. Mohar}} & & 125.00 & Empty & ComNTP &\\\midrule
    34 & Osservatorio di Foligno & Umbria & 42$^\circ$ 57$^\prime$ 36.972$^{\prime\prime}$ N& 30.0 cm& IMG & Negative \\
    &  & & 12$^\circ$ 42$^\prime$ 17.928$^{\prime\prime}$ E & QHY174M & 1\,s & 0.1\,s\\
    & \multicolumn{1}{l}{\parbox[t]{5.5cm}{\raggedright\em Roberto Nesci}} & & 220.00 & Empty & ComNTP &\\\midrule
    35 & Ad{\i}yaman University Astrophysics & T\"urkiye & 37$^\circ$ 45$^\prime$ 6.05999$^{\prime\prime}$ N& 61.0 cm& IMG & Negative \\
    & Application and Research Center (ADYU60) & & 38$^\circ$ 13$^\prime$ 31.188$^{\prime\prime}$ E & iKon-M 934 & 1\,s & 0.05\,s\\
    & \multicolumn{1}{l}{\parbox[t]{5.5cm}{\raggedright\em E. Sonbas, W. Og?oza}} & & 690.00 & Empty & CamGPS &\\\midrule
    36 & T\"{U}RKSAT & T\"urkiye & 39$^\circ$ 38$^\prime$ 11.8752$^{\prime\prime}$ N& 50.0 cm& IMG & Negative \\
    &  & & 32$^\circ$ 48$^\prime$ 14.9652$^{\prime\prime}$ E & Kepler KL4040 & 1\,s & 1.0\,s\\
    & \multicolumn{1}{l}{\parbox[t]{5.5cm}{\raggedright\em M. N. Bagiran}} & & 950.00 & Clear & ComNTP &\\\midrule
    37 & UZAYMER - Çukurova University & T\"urkiye & 37$^\circ$ 3$^\prime$ 20.088$^{\prime\prime}$ N& 50.0 cm& IMG & Negative \\
    &  & & 35$^\circ$ 20$^\prime$ 52.71$^{\prime\prime}$ E & FLI Proline PL16803 & 5\,s & 0\,s\\
    & \multicolumn{1}{l}{\parbox[t]{5.5cm}{\raggedright\em A.Solmaz , M.Teke? }} & & 126.00 & Clear & CamGPS &\\\midrule
    38 & Eski\c{s}ehir Technical University & T\"urkiye & 39$^\circ$ 53$^\prime$ 7.6992$^{\prime\prime}$ N& 40.0 cm& IMG & Negative \\
    & Observatory & & 30$^\circ$ 27$^\prime$ 38.4804$^{\prime\prime}$ E & FLI Proline & 10\,s & 0\,s\\
    & \multicolumn{1}{l}{\parbox[t]{5.5cm}{\raggedright\em Metin Altan}} & & 1005.00 & Clear & ComGPS &\\\midrule
    39 & Piszkéstet? & Hungary & 47$^\circ$ 55$^\prime$ 1.5235$^{\prime\prime}$ N& 100.0 cm& IMG & Overcast \\
    &  & & 19$^\circ$ 53$^\prime$ 41.83$^{\prime\prime}$ E & Andor iXon+888 & -\ & 0.01\,s\\
    & \multicolumn{1}{l}{\parbox[t]{5.5cm}{\raggedright\em Nora Takacs, Zsófia Bora}} & & 942.67 & E & ComGPS &\\\midrule
    40 & Astronomical Station & Serbia & 43$^\circ$ 8$^\prime$ 24.8388$^{\prime\prime}$ N& 140.0 cm& IMG & Overcast \\
    & Vidojevica & & 21$^\circ$ 33$^\prime$ 20.088$^{\prime\prime}$ E & Andor IkonL & -\ & -\\\
    & \multicolumn{1}{l}{\parbox[t]{5.5cm}{\raggedright\em D. Ilic, M. Grozdanovic}} & & 1145.22 & L & CamGPS &\\\midrule
    41 & Stazione Osservativa di Cima & Italy & 45$^\circ$ 50$^\prime$ 55.3252$^{\prime\prime}$ N& 91.0 cm& IMG & Overcast \\
    & Ekar & & 11$^\circ$ 34$^\prime$ 8.63234$^{\prime\prime}$ E & KAF-16803 & -\ & 30.0\,s\\
    & \multicolumn{1}{l}{\parbox[t]{5.5cm}{\raggedright\em D. Nardiello, V. Nascimbeni}} & & 1369.90 & Empty & ComNTP &\\\midrule
    42 & Nandrin SAL & Belgium & 50$^\circ$ 31$^\prime$ 24.9312$^{\prime\prime}$ N& 40.6 cm& VID & Overcast \\
    &  & & 5$^\circ$ 26$^\prime$ 29.3604$^{\prime\prime}$ E & Watec 910 HX /RC & -\ & 0\,s\\
    & \multicolumn{1}{l}{\parbox[t]{5.5cm}{\raggedright\em Olivier Schreurs, Manon Lecossois}} & & 261.00 & Clear & Other &\\\midrule
    43 & University of Athens Observatory & Greece & 37$^\circ$ 58$^\prime$ 6.8196$^{\prime\prime}$ N& 40.0 cm& IMG & Overcast \\
    & (UOAO) & & 23$^\circ$ 47$^\prime$ 0.1248$^{\prime\prime}$ E & SBIG ST10 XME & -\ & 3\,s\\
    & \multicolumn{1}{l}{\parbox[t]{5.5cm}{\raggedright\em Kosmas Gazeas}} & & 250.00 & Empty & ComNTP &\\\midrule
    44 & Galhassin Robotic Telescope, & Italy & 37$^\circ$ 56$^\prime$ 21.6996$^{\prime\prime}$ N& 40.0 cm& IMG & Overcast \\
    & Isnello & & 14$^\circ$ 1$^\prime$ 14.196$^{\prime\prime}$ E & FLI ProLine 16803 & -\ & 1.5\,s\\
    & \multicolumn{1}{l}{\parbox[t]{5.5cm}{\raggedright\em A. Nastasi}} & & 649.49 & Luminance & ComGPS &\\\midrule
    45 & Ondokuz May{\i}s University & T\"urkiye & 41$^\circ$ 22$^\prime$ 3.8172$^{\prime\prime}$ N& 37.0 cm& IMG & Overcast \\
    & Observatory & & 36$^\circ$ 12$^\prime$ 5.6736$^{\prime\prime}$ E & SBIG STL-4020M CCD & -\ & 9.8\,s\\
    & \multicolumn{1}{l}{\parbox[t]{5.5cm}{\raggedright\em S. Kalkan}} & & 150.00 & Empty & ComGPS &\\\midrule
    46 & CNVL Observatory, Baia Mare & Romania & 47$^\circ$ 39$^\prime$ 28.8242$^{\prime\prime}$ N& 25.4 cm& IMG & Overcast \\
    &  & & 23$^\circ$ 34$^\prime$ 7.55636$^{\prime\prime}$ E & Atik 383L+ & -\ & 4.0\,s\\
    & \multicolumn{1}{l}{\parbox[t]{5.5cm}{\raggedright\em L. Stoian}} & & 270.69 & Empty & ComNTP &\\\midrule
    47 & Institute of Space Science & Romania & 44$^\circ$ 21$^\prime$ 5.11591e-12$^{\prime\prime}$ N& 20.3 cm& IMG & Overcast \\
    &  & & 26$^\circ$ 1$^\prime$ 48$^{\prime\prime}$ E & Meade DSI III Pro & -\ & 0.5\,s\\
    & \multicolumn{1}{l}{\parbox[t]{5.5cm}{\raggedright\em Dumitru Bogdan Alexandru}} & & 90.00 & Empty & ComNTP &\\\midrule
    48 & Home Legionowo & Poland & 52$^\circ$ 22$^\prime$ 57.2714$^{\prime\prime}$ N& 20.3 cm& IMG & Overcast \\
    &  & & 20$^\circ$ 54$^\prime$ 8.97207$^{\prime\prime}$ E & ZWO ASI120mm & -\ & 0.1\,s\\
    & \multicolumn{1}{l}{\parbox[t]{5.5cm}{\raggedright\em Daniel Antuszewicz}} & & 80.08 & Empty & ComNTP &\\\midrule
    49 & Zeta Aquarii Observatory - Mobile & Romania & 45$^\circ$ 38$^\prime$ 0.800016$^{\prime\prime}$ N& 20.0 cm& IMG & Overcast \\
    & Station & & 25$^\circ$ 33$^\prime$ 49.28$^{\prime\prime}$ E & Canon 1100D & -\ & 2.0\,s\\
    & \multicolumn{1}{l}{\parbox[t]{5.5cm}{\raggedright\em Lucian Curelaru, Sorin Ion}} & & 812.00 & Clear & CamGPS &\\\midrule
    50 & \.{I}STEK Belde Observatory & T\"urkiye & 41$^\circ$ 1$^\prime$ 49.0476$^{\prime\prime}$ N& 10.0 cm& IMG & Overcast \\
    &  & & 29$^\circ$ 2$^\prime$ 33.3996$^{\prime\prime}$ E & QHY 5 III 178 M & -\ & 0.0\,s\\
    & \multicolumn{1}{l}{\parbox[t]{5.5cm}{\raggedright\em M.Acar }} & & 110.00 & Empty & ComNTP &\\\midrule\midrule
\end{longtable}
\end{small}       

\clearpage

\end{appendix}

\end{document}